\documentclass[aip,preprint]{revtex4-2}
\usepackage{tikz}
\usepackage{pgfplots}
\usepackage{pgfplotstable}
\pgfplotsset{width=10cm,compat=1.9}
\usepackage[cp1251]{inputenc}
\usepackage[T2A]{fontenc}
\usepackage[english]{babel}
\usepackage{amssymb,latexsym,amsmath,amscd}
\usepackage{graphicx,color,framed}
\usepackage{bm}
\usepackage{appendix}
\usepackage{tabu}
\usepackage{tikz}
\usepackage{xcolor}
\usepackage{siunitx}
\usepackage{empheq}
\usepackage{float}
\usepackage{subcaption}
\usepackage{xr}
\usepackage{siunitx}
\usepackage{microtype}
\usepackage{amsfonts}
\usepackage{gensymb}
\usepackage{hyperref}
\usepackage[normalem]{ulem}
\graphicspath{{figures/}}

\makeatletter
\newcommand*{\addFileDependency}[1]{
  \typeout{(#1)}
  \@addtofilelist{#1}
  \IfFileExists{#1}{}{\typeout{No file #1.}}
}
\makeatother

\begin{document}

\author{\firstname{Yury A.} \surname{Budkov}}
\email[]{ybudkov@hse.ru}
\affiliation{Laboratory of Computational Physics, HSE University, Tallinskaya st. 34, 123458 Moscow, Russia}
\affiliation{School of Applied Mathematics, HSE University, Tallinskaya st. 34, 123458 Moscow, Russia}
\affiliation{A.N. Frumkin Institute of Physical Chemistry and Electrochemistry, Russian Academy of Sciences, 119071, 31 Leninsky Prospect,
Moscow, Russia}
\author{\firstname{Nikolai N.} \surname{Kalikin}}
\affiliation{Laboratory of Computational Physics, HSE University, Tallinskaya st. 34, 123458 Moscow, Russia}
\affiliation{Laboratory of Multiscale Modeling of Molecular Systems, G.A. Krestov Institute of Solution Chemistry of the Russian Academy of Sciences, 153045, Akademicheskaya st. 1, Ivanovo, Russia}
\author{\firstname{Petr E.} \surname{Brandyshev}}
\affiliation{Laboratory of Computational Physics, HSE University, Tallinskaya st. 34, 123458 Moscow, Russia}
\affiliation{School of Applied Mathematics, HSE University, Tallinskaya st. 34, 123458 Moscow, Russia}
\title{A Fluctuation Theory of Liquid-Phase Solutions: Shear Viscosity}

\begin{abstract}
Accurately describing liquids and their mixtures beyond equilibrium remains a significant challenge in modern chemical physics and physical chemistry, especially regarding the calculation of transport properties in liquid-phase systems. This paper introduces a phenomenological nonequilibrium theory specifically designed for multicomponent liquid-phase solutions. Our field-theoretical framework, rooted in nonequilibrium statistical mechanics, incorporates quasi-stationary concentration fluctuations that align with equilibrium liquid theory as described by classical density functional theory. This method serves as a phenomenological extension of the established Dean-Kawasaki stochastic density functional theory, enabling the computation of shear viscosity. We apply our approach to derive general formula for the shear viscosity in single-solute solutions. Our findings yield new results and successfully reproduce previously established results for such systems as solutions containing soft-core particles, hard spheres, one-component plasma, and near-critical solutions.
\end{abstract}
\maketitle

\section{Introduction}

In the realm of contemporary chemical physics and physical chemistry, one of the most pressing challenges lies in the comprehensive characterization of liquids and their solutions when they are pushed beyond equilibrium. This challenge is particularly crucial for accurately determining the transport properties of liquid-phase systems, which indispensable in a multitude of scientific and industrial applications. Despite notable advancements in this field~\cite{de2023perspective,mills2024dynamic,illien2024dean}, a unified statistical theory capable of robustly describing non-equilibrium processes remains elusive. Such a theory would ideally facilitate direct calculations of transport properties in complex liquids. In stark contrast, Gibbs' equilibrium statistical mechanics provides a well-established framework for determining fundamental thermodynamic and structural properties of liquids, elucidating the intricacies of phase behavior and molecular interactions~\cite{hansen2013theory,barrat2003basic}. The lack of a parallel theoretical framework for non-equilibrium scenarios hinders our understanding of liquid-state dynamics, revealing a critical gap in current scientific literature and emphasizing the need for further research in this essential area.

The dynamical density functional theory (DDFT) is a promising candidate for describing fluid dynamics~\cite{mills2024dynamic,de2023perspective,rex2008dynamical,donev2014dynamic}, but it has some limitations. This theory, formulated in terms of dynamic equations for the average density of a fluid, does not directly calculate the transport properties of fluids, such as diffusion, viscosity, and electrical conductivity. The exception is a recent study~\cite{aerov2015theory} that attempted to calculate the effective viscosity of hard spheres confined in a slit-like pore.

In recent years, significant progress in this direction has been made in the application of Dean-Kawasaki stochastic density functional theory (SDFT) \cite{dean1996langevin, demery2016conductivity, avni2022conductivity, avni2022conductance, robin2024correlation, illien2024dean}. This approach is rooted in the derivation of Langevin-type stochastic equations that describe the concentrations of fluid species, originating from the stochastic inertia-free (overdamped) Langevin dynamics of interacting fluid molecules~\cite{zwanzig2001nonequilibrium}. Although SDFT theoretically enables the calculation of transport properties, it is subject to inherent physical constraints. Specifically, SDFT equations arise from Langevin dynamics, which are rigorously valid for sufficiently rarefied systems where many-body correlations can be effectively represented through a white noise heat bath. This limitation is highlighted by the fact that SDFT produces a diagonal mobility matrix, expressed as $L_{ab} = m_a c_a \delta_{ab}$ (where $c_a$ denotes the concentration of species $a$), a scenario that typically only occurs in relatively dilute liquid-phase solutions~\cite{akaberian2023nonequilibrium}. Furthermore, the kinetic equations governing the average concentrations of the components correspond to DDFT equations in the inertia-free (diffusion) regime~\cite{mills2024dynamic,de2023perspective}, especially when considering the free energy functional of a liquid within the framework of random phase approximation (RPA)~\cite{illien2024dean, archer2001binary}, where the direct correlation functions are approximated as $c_{ab}(\bold{r}) = -V_{ab}(\bold{r}) / k_B T$, with $V_{ab}(\bold{r})$ representing the pair potentials~\cite{budkov2024statistical}.

Given these limitations, we argue that a generalization of SDFT is warranted, particularly for strongly correlated fluids. Such a generalization would align with the principles of equilibrium liquid-state statistical theory. However, the persistent challenges in establishing a microscopic foundation for such a theory within nonequilibrium statistical mechanics suggest that a purely first-principles approach may currently be intractable. To address these challenges, we propose developing a theory rooted in phenomenological framework of nonequilibrium thermodynamics~\cite{kondepudi2014modern,klimontovich2024statistical}, augmented by the theory of quasi-stationary fluctuations (see refs.~\cite{landau2013statistical} and Appendix \ref{quasi-stationary_fluctuations_app}). While this phenomenological approach may limit our capacity for a detailed microscopic description of fluid kinetics~\cite{akaberian2023nonequilibrium}, it enables the coupling of the resulting stochastic equations for species concentrations with classical equilibrium liquid theory, notably through a general form of classical density functional theory (cDFT)~\cite{evans2009density, sammuller2025neural}.

Essentially, this work marks an initial step towards the development of a phenomenological theory designed specifically for multicomponent liquid-phase solutions. Our framework is anchored in the principles of nonequilibrium statistical mechanics, while naturally incorporating solute concentration fluctuations consistent with equilibrium liquid theory, as captured through DFT. This approach makes it easier to calculate the shear viscosity in solutions with a single solute, using different models.

This paper is organized in the following manner. In Section \ref{Fluctuation_theory_section}, we briefly discuss a phenomenological extension of the Dean-Kawasaki stochastic DFT for liquid-phase solutions under external potential fields and flow -- a fluctuation theory. In Section \ref{Virial_shear_viscosity_section}, we apply the fluctuation theory to calculate the virial contribution to the shear viscosity of a single-solute solutions. Section \ref{models_applications_section} discusses applications of the derived general formula for the shear viscosity to different model systems. Section \ref{discussion_section} contains discussions on further applications of the fluctuation theory, while Section \ref{conslusion_section} provides concluding remarks. Appendices \ref{quasi-stationary_fluctuations_app}-\ref{eq_delta_eta_app} include technical details that were omitted from the main text for better readability.

\section{Fluctuation theory}\label{Fluctuation_theory_section}
Let us consider a multi-component solution consisting of particles, each representing a distinct type of molecule. These particles interact through pairwise effective potentials, denoted as $V_{ab}(|\bold{r}|)$, where $\bold{r}$ is the vector representing the interparticle distance. We model the solvent as a continuous fluid whose macroscopic dynamics obeys the hydrodynamic laws, with flow behavior characterized by a velocity field $\bold{v}(\bold{r},t)$. In the presence of an external potential field(s) with potential energies $\varphi_a(\bold{r})$, the chemical potential near the equilibrium state has the form
\begin{equation}
\mu_{a}^{(\text{tot})}=\mu_{a}(\{c_l\})+\varphi_{a}.
\end{equation}
Thereby, the total flux of the solute $a$ is~\cite{kondepudi2014modern}
\begin{equation}
\bold{J}_a=-L_{ab}\nabla\mu_{b}^{(\text{tot})}+c_a\bold{v}+\bold{j}_a,
\end{equation}
where $L_{ab}$ is the mobility matrix~\cite{akaberian2023nonequilibrium} whose elements are the functions of the local concentrations of solutes, $c_{a}$. Note that for repeated indices summation is implied. We assume that the external fields are stationary, as we will be considering a steady state which the system approaches at $t \to \infty$. The random fluxes, $\bold{j}_a(\bold{r},t)$, are not related to the gradients of chemical potentials and external solvent flow, originating exclusively from thermal motion of the solute molecules (see also Appendix \ref{Particle_flux_fluctuations_app}). It is important to note that chemical potentials lack a rigorous definition in nonequilibrium states. However, for systems near equilibrium, the concept can be extended by treating the chemical potential as a functional of nonequilibrium concentration profiles. This approach is similar to the framework of cDFT, which describes the equilibrium intrinsic chemical potentials as the functionals of local concentrations of species (for a broader discussion, see Appendix \ref{Particle_flux_fluctuations_app}, technical details are provided in Appendix \ref{Equations_of_motion_Zwanzig_equation_app}).

The velocity field satisfies the following linearized Navier-Stokes equation~\cite{maduar2015electrohydrodynamics,vinogradova2023electrophoresis,avni2022conductance,avni2022conductivity}
\begin{equation}
\label{stokes}
\eta_0 \nabla^2 \bold{v} -\nabla p + \bold{G}_{\text{ext}}+\bold{G}_{\text{int}}+\boldsymbol{\xi}=0,
\end{equation}
where we have neglected the small quadratic inertial term, $\sim (\bold{v} \cdot\nabla) \bold{v}$. The latter approximation is valid for rather small velocity field~\cite{golestanian2025hydrodynamically}, as is the case of present work. $\eta_0$ denotes the shear viscosity of pure solvent. We have introduced a Gaussian white noise
term defined via~\cite{golestanian2025hydrodynamically} $\left<\boldsymbol{\xi}(\bold{r},t)\right>=0$ and
\begin{equation}
\left<\xi_{i}(\bold{r},t)\xi_{k}(\bold{r}',t')\right>=2\eta_0 k_{B}T(-\delta_{ik}\nabla^2 +\partial_{i}\partial_{k})\delta(\bold{r}-\bold{r}')\delta(t-t'),
\end{equation}
which describes the hydrodynamic fluctuations~\cite{de2006hydrodynamic,landau1992hydrodynamic,golestanian2025hydrodynamically} of solvent and respects the solvent incompressibility condition, $\nabla\cdot \bold{v}=0$.

The pressure of the solution, $p$, is equal to the sum of the solvent's pressure and the Van't Hoff's osmotic pressure from the solutes, i.e. $p = p_s + \sum_a c_a k_BT$; $\bold{G}_{\text{ext}}=\bold{f}_{a}c_a=-c_a\nabla \varphi_a$ is the external force density acting on $a$-th molecular species. For example, this force could be caused by an electrostatic field acting on charged particles, such as ions or colloids, or by a gravitational field acting on colloids. The internal force density~\cite{schmidt2022power} is
\begin{equation}
\bold{G}_{\text{int}}(\bold{r},t)=-c_a(\bold{r},t)\nabla \mu_{a,ex}(\bold{r},t)
\end{equation}
with the excess chemical potential, $\mu_{a,ex}(\bold{r},t)=-k_{B}T c_{1}^{(a)}(\bold{r},t)$, of species $a$, where $c_{1}^{(a)}(\bold{r},t)$ is the one-particle correlation function of species $a$. The latter is the functional of the local concentrations of solutes~\cite{hansen2013theory,budkov2024statistical}.

The total flux fluctuation in the linear approximation over the fluctuating variables is
\begin{equation}
\label{delta_J}
\delta \bold{J}_a=\bold{J}_a-\left<\bold{J}_a\right>=-\bar{L}_{ab} \nabla\delta\mu_b -\frac{\partial \bar{L}_{ab}}{\partial{\bar{c}_l}}\delta c_l \nabla \varphi_b +\bar{c}_a \delta\bold{v} +\delta c_a\bar{\bold{v}}+\delta \bold{j}_{a},
\end{equation}
where, $\bar{L}_{ab}={L}_{ab}(\{\bar{c}\})$, $\delta \bold{v}$ is the fluctuation of the velocity field occurred due to the fluctuations of the local concentrations of solutes, $\bar{\bold{v}}=\left<\bold{v}\right>$ -- average velocity field, $\bar{c}_a$ -- average concentration of solute $a$. The random fluxes, $\delta \bold{j}_{a}=\bold{j}_{a}-\left<\bold{j}_{a}\right>=\bold{j}_{a}$, satisfy the following relations:  
\begin{equation}
\label{fdt}
\left<\delta \bold{j}_{a}(\bold{r},t)\right>=0,\quad \left<\delta j_{a i}(\bold{r},t)\delta j_{b k}(\bold{r}',t')\right>=2\bar{L}_{ab} k_{B}T\delta_{ik}\delta(\bold{r}-\bold{r}')\delta(t-t'),
\end{equation}
where $k_B$ is the Boltzmann constant, $T$ is the temperature. Note that throughout the paper, we refer to correlations of fluctuating fluxes as being local in space and time. This means that we assume the size of the small volumes of solution we consider for fluctuations is much larger than the typical range of interaction between solute molecules. We also assume the time scale we use for diffusion is much longer than the microscopic correlation time~\cite{klimontovich2024statistical}. Note that we consider the case of sufficiently weak external fields and flow, so we can neglect their effect on the fluctuation-dissipation relation (\ref{fdt}). In other words, we can consider it equivalent to the one for the equilibrium solution (see Appendix \ref{Particle_flux_fluctuations_app}). In recent papers~\cite{cui2018generalized,pelargonio2023generalized} Zaccone and coauthors addressed instructive microscopic considerations, where these effects need to be taken into account.

Taking into account solution incompressibility, which constraints both the average velocity field and its fluctuation, and using continuity equation (see also eq. (\ref{discontin}))
\begin{equation}
\partial_t\delta {c_a} = -\nabla \cdot\delta\bold{J}_a,
\end{equation}
we can obtain the following equations of motion
\begin{equation}
\label{stoch_eq}
\partial_t \delta c_a = -\hat{\mathcal{L}}_{ab}\delta c_{b} +\delta \eta_a,
\end{equation}
where the random noises, $\delta\eta_a=-\nabla\cdot\delta \bold{j}_a$, satisfy the following fluctuation-dissipation relation
\begin{equation}
\label{correlator___}
\left<\delta\eta_{a}(\bold{r},t)\delta\eta_{b }(\bold{r}',t')\right>=-2\bar{L}_{ab} k_B T\nabla^2\delta(\bold{r}-\bold{r}')\delta(t-t'),
\end{equation}
and $\hat{\mathcal{L}}_{ab}$ is the Liouvillian, represented by:
\begin{equation}
\label{evolution_}
\hat{\mathcal{L}}_{ab}=\hat{\mathcal{L}}_{ab}^{(\text{eq})}+\hat{\mathcal{L}}_{ab}^{(\text{ex})}+\hat{\mathcal{L}}_{ab}^{(\text{ad})}.
\end{equation}
The first term on the right-hand side of eq. (\ref{evolution_}) is the "equilibrium" part of the Liouvillian which acts on the functions in accordance with the rule (see also eq. (\ref{operator}) in Appendix \ref{Equations_of_motion_Zwanzig_equation_app})
\begin{equation}
\label{operator_}
\hat{\mathcal{L}}_{ab}^{(\text{eq})}f_{b}(\bold{r})=- k_{B}T\bar{L}_{ac}\int d^{3}\bold{r}'\,\nabla^2C_{cb}(\bold{r},\bold{r}')f_{b}(\bold{r}'),
\end{equation}
where
\begin{equation}
\label{Cab__}
C_{ab}(\bold{r},\bold{r}')=\frac{1}{\bar{c}_a}\delta_{ab}\delta(\bold{r}-\bold{r}')- c_{ab}(\bold{r}-\bold{r}')
\end{equation}
with the direct correlation functions, $c_{ab}(\bold{r}-\bold{r}')$, of the species (see also eq. (\ref{Cab_}) in Appendix \ref{Equations_of_motion_Zwanzig_equation_app}). The second term is related to the effect of the external potential fields, and is given by: 
\begin{equation}
\label{evolution_2}
\hat{\mathcal{L}}_{ab}^{(\text{ex})}=-\Gamma_{abl}\left(\nabla^2\varphi_l  + \nabla\varphi_l \cdot \nabla\right), \qquad \Gamma_{abl}=\frac{\partial \bar{L}_{al}}{\partial{\bar{c}_b}}.
\end{equation}
Finally, the third term represents the macroscopic flow (advection) effect, which is:
\begin{equation}
\label{evolution_3}
\hat{\mathcal{L}}_{ab}^{(\text{ad})}=\delta_{ab}\bar{\bold{v}}\cdot \nabla.
\end{equation}
The stochastic equations of motion (\ref{stoch_eq}) with the corresponding fluctuation-dissipation relation (\ref{correlator___}) must be solved with some initial conditions $\delta c_a(\bold{r},0)=g_{a}(\bold{r})$. The formal solution is~\cite{kruger2017gaussian}
\begin{equation}
\label{formal_solution}
\delta c_a (\bold{r},t)=g_a(\bold{r})+\int\limits_{0}^{t}d\tau \left(e^{-(t-\tau)\hat{\mathcal{L}}}\right)_{ab}\delta \eta_b(\bold{r},\tau).
\end{equation}

Note that the equations of motion for concentration fluctuations, together with the Stokes equation, resemble the stationary case ($\partial_t \bold{v}=0$) of Model H in the Hohenberg-Halperin classification~\cite{hohenberg1977theory,siggia1976renormalization,chen2024critical}. This is in contrast to the equations of motion for the case of a stationary solvent ($\bold{v}=0$) in the absence of the external forces ($\nabla\varphi_a =0$), presented in Appendix \ref{Equations_of_motion_Zwanzig_equation_app}, which defines the case of Model B~\cite{kruger2017gaussian} in this classification.

Further, expanding the internal force density in terms of concentration fluctuations, we obtain in quadratic approximation
\begin{equation}
\bold{G}_{\text{int}}(\bold{r},t)\approx k_{B}T\bar{c}_a\nabla \int d^{3}\bold{r}'\,c_{ab}(\bold{r}-\bold{r}')\delta c_b(\bold{r}',t)+k_{B}T\delta c_a(\bold{r},t)\nabla \int d^{3}\bold{r}'\,c_{ab}(\bold{r}-\bold{r}')\delta c_b(\bold{r}',t),
\end{equation}
where we take into account that in the linear approximation of the concentration fluctuation
\begin{equation}
c_{1}^{(a)}(\bold{r},t)=-\beta \mu_{\text{ex,bulk}}^{(a)} + \int d^{3}\bold{r}'\,c_{ab}(\bold{r}-\bold{r}')\delta c_b(\bold{r}',t), \qquad c_{ab}(\bold{r}-\bold{r}')=\frac{\delta c_1^{(a)}(\bold{r},t)}{\delta c_b(\bold{r}',t)}\bigg{|}_{c_l=\bar{c}_l},
\end{equation}
where $\mu_{\text{ex,bulk}}^{(a)}$ is the excess chemical potential of solute $a$ in the bulk solution at equilibrium.

Now, averaging (\ref{stokes}) over the statistics of the concentration and velocity fluctuations, taking into account that $\left<\delta c_a(\bold{r},t)\right>=0$ and $\left<\boldsymbol{\xi}(\bold{r},t)\right>=0$, we obtain
\begin{equation}
\label{stokes_2}
\eta_0\nabla^2\bar{\bold{v}} - \nabla \bar{p}+\bar{\bold{G}}_{\text{ext}}+\bar{\bold{G}}_{\text{int}}=0,
\end{equation}
where $\bar{p}(\bold{r},t)=\left<p(\bold{r},t)\right>$ is the average pressure, $\bar{\bold{G}}_{\text{ext}}(\bold{r},t)=\left<{\bold{G}}_{\text{ext}}(\bold{r},t)\right>$ is the average external force density, and
\begin{equation}
\bar{\bold{G}}_{\text{int}}(\bold{r},t)=\left<\bold{G}_{\text{int}}(\bold{r},t)\right>=k_{B}T \int d^{3}\bold{r}'\,\nabla c_{ab}(\bold{r}-\bold{r}')S_{ab}(\bold{r},\bold{r}',t)
\end{equation}
is the average internal force density, with $S_{ab}(\bold{r},\bold{r}',t)=\left<\delta c_{a}(\bold{r},t)\delta c_b(\bold{r}',t)\right>$ defining the correlation function (correlator) of instantaneous concentration fluctuations goes to the steady state value at $t\to \infty $, i.e. $S_{ab}(\bold{r},\bold{r}',t)\to  S_{ab}(\bold{r},\bold{r}')$, which, in turn, satisfies the Zwanzig equation (see ref. \cite{zwanzig2001nonequilibrium})
\begin{equation}
\label{zwanzig_main}
\int d^3\bold{r}^{\prime\prime}\,{\mathcal{L}}_{ac}(\bold{r},\bold{r}^{\prime\prime}) S_{cb}(\bold{r}^{\prime\prime},\bold{r}')+\int d^3\bold{r}^{\prime\prime}\,{\mathcal{L}}_{bc}(\bold{r}',\bold{r}^{\prime\prime}) S_{ca}(\bold{r}^{\prime\prime},\bold{r}) = -2 k_{B}T\bar{L}_{ab}\nabla^2\delta(\bold{r}-\bold{r}'),
\end{equation}
where ${\mathcal{L}}_{ab}(\bold{r},\bold{r}^{\prime})=\hat{\mathcal{L}}_{ab}\delta(\bold{r}-\bold{r}^{\prime})$ is the kernel of Liouvillian (see also Appendix \ref{Equations_of_motion_Zwanzig_equation_app}). Thus, at $t\to \infty$, the averaged internal force density in the Fourier-representation is
\begin{equation}
\label{volume_force_}
\bar{\bold{G}}_{\text{int}}(\bold{k})=k_{B}T \int\frac{d^3 \bold{q}}{(2\pi)^3} i\bold{q}c_{ab}(\bold{q})S_{ab}(\bold{k}-\bold{q},\bold{q}).
\end{equation}
At $\bold{k}\to 0$ the Fourier-image of the internal force density should behave as~\cite{robin2024correlation} $\bar{\bold{G}}_{\text{int}}(\bold{k})\simeq -\Delta \eta k^2 \bar{\bold{v}}(\bold{k})$ leading to the following linearized Navier-Stokes equation at large length scale 
\begin{equation}
(\eta_0 +\Delta\eta) \nabla^2 \bar{\bold{v}}(\bold{r}) -\nabla  \bar{p}(\bold{r})+\bar{\bold{G}}_{\text{ext}}(\bold{r})=0,
\end{equation}
so that the shear viscosity of solution is
\begin{equation}
\eta =\eta_0+\Delta\eta.
\end{equation}

Thus, to calculate the shear viscosity, we have to solve the Zwanzig equation (\ref{zwanzig_main}), calculate the Fourier transform of the correlators $S_{ab}(\bold{r},\bold{r}')$, substitute it into equation (\ref{volume_force_}), and then take the limit of small $\bold{k}$. In the next section, we discuss how this program can be applied to single-solute solutions.

It is important to note that in the general case, the kinetic description of multicomponent solutions uses an off-diagonal mobility matrix, $\bar{L}_{ab}$. However, in sufficiently dilute solutions -- where the solvent concentration is much higher than that of the solutes -- we can reasonably simplify this matrix to a diagonal form: $\bar{L}_{ab} = m_a\bar{c}_a \delta_{ab}$, where $m_a>0$ is the mobility of solute $a$ at infinite dilution. In fact, the latter approximation is the first-order term in the mobility matrix series expansion with respect to the concentrations of species. This approximation together with the approximation $c_{ab}(\bold{r})=-\beta V_{ab}(|\bold{r}|)$ is consistent with the Dean-Kawasaki SDFT~\cite{dean1996langevin,illien2024dean,demery2016conductivity}, based on an in-depth microscopic analysis of Langevin particle dynamics. In contrast to the microscopic derivations, proposed approach is based on the phenomenology of nonequilibrium thermodynamics, while incorporating the principles of both local equilibrium and classical DFT. In principle, within the framework of this formalism, it is possible to consider the off-diagonal elements of the mobility matrix on a phenomenological level. The first terms in the concentration expansion of the off-diagonal elements are anticipated to be proportional to the product of the concentrations of the components~\cite{akaberian2023nonequilibrium}. Specifically, this can be expressed as $L_{ab} = \alpha_{ab} c_a c_b$ for $a \neq b$, where $\alpha_{ab}$ denotes phenomenological symmetric coefficients that cannot be determined within the current thermodynamic formalism. Note that the calculation of mobility matrix elements is a complex problem that should be addressed within the framework of physical kinetics~\cite{pitaevskii2012physical} or obtained from molecular simulations~\cite{akaberian2023nonequilibrium}. In this theory, the mobility matrix is purely phenomenological. In what follows we restrict our analysis only to the case of the single-solute solutions.

\section{Virial shear viscosity of single-solute solutions}\label{Virial_shear_viscosity_section}

Let us consider the case of a single solute in a solution. We assume that the flow of the solution has a local quasi-stationary velocity field, denoted by $\bold{v}(\bold{r},t)$, which satisfies the aforementioned linearized Navier-Stokes equation:
\begin{equation}
\label{Navier}
\eta_0 \nabla^2 \bold{v} - \nabla p + {\bold{G}_{\text{int}}}+\boldsymbol{\xi}= 0,
\end{equation}
where we assumed the absence of the external volume forces ($\bold{G}_{\text{ext}}=0$). In the Fourier-representation, at $t\to \infty$, the averaged linearized Navier-Stokes equation (\ref{Navier}) gives
\begin{equation}
-\eta_0 k^2 \bar{\bold{v}}(\bold{k}) -i\bold{k} \bar{p}(\bold{k})+\bar{\bold{G}}_{\text{int}}(\bold{k})=0,
\end{equation}
where
\begin{equation}
\label{volume_force}
\bar{\bold{G}}_{\text{int}}(\bold{k})=k_{B}T \int\frac{d^3 \bold{q}}{(2\pi)^3} i\bold{q}c_2(\bold{q})S(\bold{k}-\bold{q},\bold{q}),
\end{equation}
where $c_{2}(\bold{q})$ is the Fourier transform of the solute-solute direct correlation function.

As discussed in Section \ref{Fluctuation_theory_section}, to calculate the shear viscosity, we have to calculate the Fourier-transform of the correlator $S(\bold{r},\bold{r}')$, substitute it into equation (\ref{volume_force}), and then take the limit of small $\bold{k}$. For this purpose, let us write the Zwanzig equation
\begin{equation}
\label{zwanzig__}
\int d^3\bold{r}^{\prime\prime}\,\mathcal{L}(\bold{r},\bold{r}^{\prime\prime}) S(\bold{r}^{\prime\prime},\bold{r}')+\int d^3\bold{r}^{\prime\prime}\,{\mathcal{L}}(\bold{r}',\bold{r}^{\prime\prime}) S(\bold{r}^{\prime\prime},\bold{r}) = -2 k_{B}T\bar{L}\nabla^2\delta(\bold{r}-\bold{r}'),
\end{equation}
where 
\begin{equation}
\mathcal{L}(\bold{r},\bold{r}^{\prime})={\mathcal{L}}^{(\text{eq})}(\bold{r},\bold{r}^{\prime})+{\mathcal{L}}^{(\text{ad})}(\bold{r},\bold{r}^{\prime}),
\end{equation}
\begin{equation}
\mathcal{L}^{(\text{eq})}(\bold{r},\bold{r}^{\prime})=- k_{B}T\bar{L}\nabla^2C(\bold{r}-\bold{r}'), \qquad \mathcal{L}^{(\text{ad})}(\bold{r},\bold{r}^{\prime})=\bar{\bold{v}}(\bold{r})\cdot \nabla\delta(\bold{r}-\bold{r}^{\prime})
\end{equation}
or in the Fourier-representation
\begin{equation}
\label{zwanzig__Fourier}
\int \frac{d^3\bold{p}}{(2\pi)^3}\mathcal{L}(\bold{k},-\bold{p}) S(\bold{p},\bold{k}')+\int \frac{d^3\bold{p}}{(2\pi)^3}{\mathcal{L}}(\bold{k}',-\bold{p}) S(\bold{p},\bold{k}) = 2 k_{B}T\bar{L}k^2(2\pi)^3\delta(\bold{k}+\bold{k}'),
\end{equation}
\begin{equation}
\mathcal{L}(\bold{k},\bold{k}^{\prime})={\mathcal{L}}^{(\text{eq})}(\bold{k},\bold{k}^{\prime})+{\mathcal{L}}^{(\text{ad})}(\bold{k},\bold{k}^{\prime}),
\end{equation}
\begin{equation}
{\mathcal{L}}^{(\text{eq})}(\bold{k},\bold{k}^{\prime})=\bar{L}k_{B}T k^2C(\bold{k})(2\pi)^3\delta(\bold{k}+\bold{k}'), \qquad {\mathcal{L}}^{(\text{ad})}(\bold{k},\bold{k}^{\prime})=-i\bold{k}'\cdot \bar{\bold{v}}(\bold{k}+\bold{k}').
\end{equation}
Let us solve eq. (\ref{zwanzig__Fourier}) using the successive approximations method, limited by a linear term in the velocity field, i.e. we look for solution in the form
\begin{equation}
\label{S_tot}
S(\bold{k},\bold{k}')=S^{(0)}(\bold{k},\bold{k}')+S^{(1)}(\bold{k},\bold{k}'),
\end{equation}
where $S^{(0)}(\bold{k},\bold{k}')=(2\pi)^3\delta(\bold{k}+\bold{k}')S(\bold{k})$ is the solution of the Zwanzig equation at the equilibrium (see Appendix \ref{Equations_of_motion_Zwanzig_equation_app}), i.e. at $\bar{\bold{v}}=0$; $S(\bold{k})$ is the structure factor of solute; $S^{(1)}(\bold{k},\bold{k}')$ is the first-order correction to the equilibrium correlation function, which is linear in the average velocity field. After some algebra, we obtain 
\begin{equation}
\label{S1}
S^{(1)}(\bold{k},\bold{k}')=\frac{i\bold{k}'\cdot \bar{\bold{v}}(\bold{k}+\bold{k}')(S(\bold{k}')-S(\bold{k}))}{\bar{L}k_{B}T\left(\bold{k}^2 C(\bold{k})+\bold{k}'^2C(\bold{k}')\right)},
\end{equation}
where $C(\bold{k})=1/S(\bold{k})$.

As can be seen, due to the solution incompressibility condition, $\bold{k}'\cdot \bar{\bold{v}}(\bold{k}+\bold{k}')=-\bold{k}\cdot \bar{\bold{v}}(\bold{k}+\bold{k}')$, the function $S^{(1)}(\bold{k},\bold{k}')=S^{(1)}(\bold{k}',\bold{k})$. Substituting (\ref{S_tot}) into (\ref{volume_force}) and taking into account (\ref{S1}), we can obtain
\begin{equation}
\label{viscosity_dispersion}
\bar{\bold{G}}_{\text{int}}(\bold{k})=\frac{1}{\bar{L}}\int\frac{d^3\bold{q}}{(2\pi)^3}\frac{(S(\bold{q})-S(\bold{k}-\bold{q}))S(\bold{q})S(\bold{k}-\bold{q})}{(\bold{k}-\bold{q})^2S(\bold{q})+\bold{q}^2S(\bold{k}-\bold{q})}c_2(\bold{q})(\bold{q}\otimes\bold{q}):\bar{\bold{v}}(\bold{k}),
\end{equation}
where $\bold{q}\otimes\bold{q}$ is the dyadic tensor with components $q_i q_j$ and $:$ denotes the tensor convolution. Note that in accordance with the definition $\bar{\bold{G}}_{\text{int}}(\bold{k})=-\bold{k}^2 \Delta\boldsymbol{\eta}(\bold{k}):\bar{\bold{v}}(\bold{k})$, eq. (\ref{viscosity_dispersion}) determines the spatial dispersion of the viscosity tensor~\cite{kawasaki1966correlation}:
\begin{equation}
\Delta\boldsymbol{\eta}(\bold{k})=\frac{1}{\bar{L}}\int\frac{d^3\bold{q}}{(2\pi)^3}\frac{(S(\bold{k}-\bold{q})-S(\bold{q}))S(\bold{q})S(\bold{k}-\bold{q})c_2(\bold{q})}{\bold{k}^2((\bold{k}-\bold{q})^2S(\bold{q})+\bold{q}^2S(\bold{k}-\bold{q}))}\bold{q}\otimes\bold{q},
\end{equation}
which should transform into a scalar tensor at macroscopic length scale, i.e. $\Delta \eta _{ik} \simeq \Delta \eta \delta_{ik}$ at $\bold{k}\to 0$. Indeed, taking into account that $S(\bold{k})$ and $c_2(\bold{k})$ depend only on modulus of $\bold{k}$, we obtain at $\bold{k}\to 0$ (see Appendix \ref{eq_delta_eta_app})
\begin{equation}
\bar{\bold{G}}_{\text{int}}(\bold{k})\simeq -k^2 \bar{\bold{v}}(\bold{k}) \frac{\bar{c}^3}{120\pi^2 \bar{L}}\int\limits_{0}^{\infty}\frac{dq \,q^2h_2^{\prime}(q)^2}{1+\bar{c}h_2(q)}.
\end{equation}
From the latter, we can derive the following general expression for the contribution of solute-solute interactions to the solution viscosity, which we refer to as {\sl virial} shear viscosity:
\begin{equation}
\label{delta_eta}
\Delta \eta =\frac{\bar{c}^3}{120\pi^2 \bar{L}}\int\limits_{0}^{\infty}\frac{dq\,q^2h_2^{\prime}(q)^2}{1+\bar{c}h_2(q)}.
\end{equation}
We refer to this value as virial because it is related to the intermolecular interaction between solute molecules. Using the Ornstein-Zernike relation~\cite{hansen2013theory}, $h_2(k)=c_2(k)(1-\bar{c}c_{2}(k))^{-1}$, for single-solute case (refer also to eq. (\ref{OZ})), we can rewrite eq. (\ref{delta_eta}) in terms of the direct correlation function
\begin{equation}
\label{delta_eta__2}
\Delta \eta =\frac{\bar{c}^3}{120\pi^2 \bar{L}}\int\limits_{0}^{\infty}\frac{dq\,q^2 c_2^{\prime}(q)^2}{(1-\bar{c}c_2(q))^3}.
\end{equation}
Eq. (\ref{delta_eta}) (or eq. (\ref{delta_eta__2})) is the main result of this paper.

\section{Verification of eq. (\ref{delta_eta})}\label{models_applications_section}

Now, to verify eq. (\ref{delta_eta}) and obtain some new results, we apply it to the calculation of the virial viscosity in different model systems. Our analysis focuses on four representative systems: solutions of particles with soft-core potentials, hard spheres, one-component plasma (OCP), and near-critical solutions.

\textbf{Solutions of particles with soft-core pair potentials. -- } For the case of solutions of particles with soft-core potentials like Gaussian-core potential~\cite{stillinger1978study,archer2001binary,likos2001effective,budkov2019statistical} in the random phase approximation (RPA), where $c_2(q)=-\beta \tilde{V}(q)$ and $\bar{L}=m\bar{c}$, we obtain
\begin{equation}
\label{delta_eta_2}
\Delta \eta =\frac{\beta^2\bar{c}^2}{120\pi^2 m}\int\limits_{0}^{\infty}\frac{dq\,q^2\tilde{V}^{\prime}(q)^2 }{(1+\beta\bar{c}\tilde{V}(q))^3}.
\end{equation}
Note that eq. (\ref{delta_eta_2}), which is a new result of this paper, corresponds to the Dean-Kawasaki SDFT approximation.

Note also that in rather dilute solutions, where the solute-solute interactions are relatively weak, we obtain the first term of the virial expansion, which was first obtained by Kr\"{u}ger et al.~\cite{kruger2018stresses} using the Green-Kubo formula within the Gaussian field theory~\cite{kruger2017gaussian}
\begin{equation}
\Delta \eta =\frac{\beta^2\bar{c}^2}{120\pi^2 m}\int\limits_{0}^{\infty}{dq\,q^2\tilde{V}^{\prime}(q)^2}.
\end{equation}

\textbf{Hard spheres. -- } Let us apply formula (\ref{delta_eta}) to the case of hard sphere solute molecules. This could be realized in suspensions of spherical colloidal particles. For the sake of simplicity, let us consider the case of low concentrations, in order to compare it with estimates available in the literature~\cite{brady1997microstructure,russelschowalter}. In this case, for the virial viscosity we have 
\begin{equation}
\label{delta_hard_spheres}
\Delta \eta =\frac{\bar{c}^2}{120\pi^2 m}\int\limits_{0}^{\infty} dq\,q^2h_2^{\prime}(q)^2,
\end{equation}
where we assume that $\bar{L}\simeq m\bar{c}$; $m = {1}/(6\pi \eta_0 R)$ is the mobility at infinite dilution, with $R$ representing the hydrodynamic radius of a hard sphere. For relatively small concentrations, the correlation function can be approximated by Mayer function:

\begin{equation} 
\label{h_HS} 
h_2(r) \simeq f(r) =e^{-\beta V_{hc}(r)}-1 =
\begin{cases} 
-1, & \text{if}\quad r\leq d,\\ 
0 ,& \text{if}\quad r>d,
\end{cases}
\end{equation}
where $V_{hc}(r)$ is the hard-core potential with the hard-core diameter, $d$.

After performing some algebra, taking into account that 
\begin{equation}
h_2'(q) =\frac{4\pi}{q^4}\left(3(\sin(q d) - qd\cos(qd)) - (qd)^2\sin(qd)\right)
\end{equation}
after integration in (\ref{delta_hard_spheres}) we obtain the virial viscosity:
\begin{equation} 
\Delta \eta_{HS} = \frac{\pi d^5\bar{c}^2}{75 m} = \frac{36}{25} \left(\frac{2R}{d}\right)\phi_d^2 \eta_0,
\end{equation}
where we define the volume fraction of the hard spheres as $\phi_d = {\pi d^3\bar{c}}/{6}$. Our estimate for the virial viscosity falls between the estimate provided by Brady and Morris for low P\'eclet numbers in simple shear flow~\cite{brady1997microstructure}, $\Delta \eta_{HS} = \frac{6}{5} \left( \frac{2R}{d} \right) \phi_d^2 \eta_0$, and the one by Russel, Saville, and Schowalter, which also applies at low P\'eclet numbers but in extensional flow~\cite{russelschowalter}, $\Delta \eta_{HS} = \frac{12}{5} \left( \frac{2R}{d} \right) \phi_d^2 \eta_0$.

Note that the radius of particles involved in excluded volume interactions and the hydrodynamic radius can be of a similar order of magnitude, but they may not necessarily be the same~\cite{aerov2015theory,brady1997microstructure}.

Note also that the total viscosity of a rather dilute suspension of hard spheres includes an additional term $\Delta \eta_{E}=5/2\eta_0 \phi_R$ ($\phi_R =4\pi R^3\bar{c}/3$), first derived by Einstein~\cite{bird2002phenomena}, which accounts for the distortion of the velocity field around the hard spheres. This term, which is purely hydrodynamic in nature, can only be obtained through fluid mechanics~\cite{landau1987fluid} and cannot be derived using the proposed thermodynamic formalism. Note that for real rather dilute colloid suspensions for which $R \sim d/2$ the virial correction to viscosity for dilute systems is always much smaller than the Einstein's hydrodynamic contribution. For rather concentrated suspensions, it is necessary to estimate the virial contribution to viscosity by using the correlation function of hard spheres within the Percus-Yevick approximation~\cite{wertheim1963exact,thiele1963equation}. However, in this case, Einstein's approximation for the hydrodynamic contribution no longer works and it also needs to be clarified~\cite{gerard2020analysis}.

\textbf{One-component plasma. -- } Let us also consider OCP, i.e. a set of point-like charged particles (ions) of charge $ze$ immersed in a compensating structureless charged background with charge density $\rho=-ze\bar{c}$ ~\cite{levin2002electrostatic,brilliantov1998accurate,khrapak2016internal}. For example, OCP can mimic the solution of immobilized charged macromolecules, which are described by a structureless charged background, with mobile counterions~\cite{budkov2015new,budkov2013surface,brilliantov1998chain}. In this case, substituting the total correlation function of OCP within the random phase approximation~\cite{budkov2015new}
\begin{equation}
\label{h_OCP}
h_2(q)=-\frac{1}{\bar{c}}\frac{\kappa^2}{q^2+\kappa^2},
\end{equation}
into eq. (\ref{delta_eta}) and taking into account that $\bar{L}=m\bar{c}$, we arrive at
\begin{equation}
\label{delta_eta_OCP}
\Delta\eta_{OCP} =\frac{\kappa}{480\pi m}
\end{equation}
Here $\kappa=(z^2e^2\bar{c}/\varepsilon k_{B}T)^{1/2}$ is the inverse Debye radius of OCP, and $\varepsilon$ is the permittivity of the medium. Note that eq. (\ref{delta_eta_OCP}) coincides with the Falkenhagen limiting law for the excess viscosity of the dilute electrolyte solutions \cite{falkenhagen1932lxii,onsager2002irreversible,robin2024correlation,fixman1962viscosity}.

\textbf{Near-critical solution. -- }  For the case of near-critical solution we substitute the Fourier-image of the total correlation function within the mean-field approximation (see, for instance, ref.~\cite{barrat2003basic})
\begin{equation}
\label{near_crit_h}
h_2(q)=\frac{1}{\bar{c}}\left(\frac{\xi}{l}\right)^2\frac{1}{1+q^2\xi^2},
\end{equation}
into eq. (\ref{delta_eta}). Thus, in the limit of $\xi \gg l$, which is always the case in the near-critical region, we obtain that
\begin{equation}
\label{delta_eta_cr}
\Delta\eta \simeq\frac{\xi}{160\pi l^2 m}.
\end{equation}
Here $l^2=2\pi\bar{c}\int_{0}^{\infty}dr\,r^4 c_2(r)$, $l$ is the so-called Debye persistence length, $\xi=l\sqrt{k_B T/(\bar{c}\partial \mu/\partial\bar{c})}$ is the correlation length.

The viscosity (\ref{delta_eta_cr}) is proportional to the correlation length that is consistent with the result first obtained by Fixman~\cite{fixman1962viscosity} and then reproduced by Kawasaki~\cite{kawasaki1966correlation} using different approaches. The shear viscosity-correlation length proportionality has been established for critical fluid within the Gaussian approximation through the Green-Kubo formula, as described in book~\cite{barrat2003basic}. Note that in order to calculate the viscosity of the near-critical solution, it is necessary to know the solute-solute direct correlation function for estimation of the Debye persistence length, $l$. However, in practice, we can use length $l$ as an adjustable parameter. In the near-critical region, this length behaves in a regular manner, while the correlation length, $\xi$, diverges as it approaches the critical point, leading to a divergence in the shear viscosity. This is because the motion of solute molecules is closely linked to the motion of other molecules within the correlation radius, $\xi$. In other words, solute molecules tend to cluster together in groups of approximately $\xi$ size. These near-critical clusters, in turn, increase the shear viscosity as the system approaches the critical point. This unusual behavior of viscosity further confirms the fact that in the vicinity of the critical point the solution is never "dilute"~\cite{gitterman1988dilute}.

\section{Discussion}\label{discussion_section}

It is essential to point out that the viscosity (\ref{delta_eta}) of a solution containing a single solute is fully determined by the intermolecular interactions of the dissolved molecules within the solvent, described by a central pair potential of interaction. Moreover, this expression does not account for the effects of hydrodynamic interactions. Accounting for the hydrodynamic interactions, which are important for colloids and macromolecules~\cite{rotne1969variational,rex2008dynamical,donev2014dynamic,henrich2007nonequilibrium,bergenholtz2001theory} is beyond the scope of the pure thermodynamic theory presented here. This has already been mentioned in the context of hard sphere applications.

Now, we would like to discuss possible future developments of the proposed theoretical framework. In our opinion, the theory presented in this paper offers a robust framework for calculation of the shear viscosity of multicomponent solutions. A noteworthy aspect of this approach is its flexibility in incorporating equilibrium correlation functions from diverse sources. Such data can be derived from computer simulations or from various approximations of integral equation theory, such as the Mean Spherical Approximation (MSA)~\cite{bernard2023analytical}.  Its applicability extends to the viscosity calculations of condensed systems characterized by strong Coulomb correlations, such as room temperature ionic liquids and concentrated electrolyte solutions. The model is particularly adaptable, incorporating not only microscopic molecular parameters of species, but also phenomenological parameters, including elements of the mobility matrix and a reference viscosity, $\eta_0$. This is a topic that will be covered in upcoming publications. In addition, the present theory can be extended to molecular liquids whose pair potential of interaction depends on the orientation of molecules, as in liquid crystals. Furthermore, the theoretical framework can be extended to account for additional effects, including thermal conductivity and thermal diffusion. By incorporating temperature fluctuations and their correlations with concentration fluctuations, as outlined in the literature~\cite{de2006hydrodynamic,kawasaki1966correlation}, we anticipate small corrections to the calculated shear viscosity. This is also a topic for future research. 

\section{Conclusions}\label{conslusion_section}

In conclusion, this study presents a phenomenological theoretical framework for multicomponent liquid-phase solutions, that bridges nonequilibrium thermodynamics and equilibrium liquid theory through classical density functional theory. By incorporating quasi-stationary fluctuations into the theory, we have calculated virial contribution to the shear viscosity of single-solute solutions. This framework not only enhances theoretical tools for modeling the nonequilibrium behavior of liquid-phase solutions, but also establishes a foundation for further research on multicomponent systems characterization.

\textbf{Data availability statement.} {\sl The data that supports the findings of this study are available within the article.}

{\bf Acknowledgements.} 
The authors would like to thank the reviewers for their valuable comments and suggestions, which have significantly improved the paper. This work is an output of a research project implemented as part of the Basic Research Program at the National Research University Higher School of Economics (HSE University). The fluctuation theory, as formulated in Appendices B and C, was partially supported by the Ministry of Science and Higher Education of the Russian Federation.

\appendix

\section{A brief reminder of the theory of quasi-stationary fluctuations}\label{quasi-stationary_fluctuations_app}

In this appendix we would like to recall some information from non-equilibrium thermodynamics and theory of quasi-stationary fluctuations. In a system, there are non-equilibrium fluxes $J_{\alpha}$ ($\alpha = 1, 2, ..., s$) that are initiated by thermodynamic forces $F_{\alpha}$. These fluxes also occur due to random fluctuations $j_{\alpha}(t)$ with zero expectation values ($\left<j_{\alpha}(t)\right>=0$), resulting in the following linear relationship:
\begin{equation}
\label{flux}
J_{\alpha} (t) = - \gamma_{\alpha\lambda} F_\lambda(t) + j_\alpha(t),
\end{equation}
where the coefficients $\gamma_{\alpha\lambda}=\gamma_{\lambda\alpha}$ are known as the Onsager coefficients\footnote{For the sake of simplicity, we assume that there is no external magnetic field present, for which the Onsager kinetic coefficients are always symmetrical.}. Recall that summation over repeated indices is assumed. The rate of entropy production is
\begin{equation}
\label{rate_production}
\dot{S}/k_B=- J_{\alpha} F_{\alpha},
\end{equation}
where $k_B$ is the Boltzmann constant, and the random fluxes have the following correlation functions~\cite{landau2013statistical,klimontovich2024statistical}
\begin{equation}
\label{fluctuation_dissipation}
\left<j_\alpha(t)j_{\lambda}(t')\right>=2\gamma_{\alpha\lambda}\delta(t-t')
\end{equation}
and zero expectation value, i.e. $\left<j_\alpha(t)\right>=0$.

The expression (\ref{fluctuation_dissipation}) plays a central role in the theory of quasi-stationary fluctuations~\cite{landau2013statistical}. In particular, Landau and Lifshitz used it to develop the theory of hydrodynamic fluctuations of the stress tensor and the heat flux~\cite{landau1992hydrodynamic}. This theoretical framework is the basis for our description of quasi-stationary fluctuations in the particle fluxes in multicomponent solutions of interacting particles.

\section{Particle flux fluctuations in the equilibrium solutions}\label{Particle_flux_fluctuations_app}

Let us consider a multi-component solution consisting of particles, each representing a distinct type of molecule. These particles interact through pairwise effective potentials, denoted as $V_{ab}(|\bold{r}|)$, where $\bold{r}$ is the vector representing the interparticle distance. For simplicity, we consider the solvent to be a continuous, non-moving liquid medium, i.e., the velocity field of the liquid is zero, $\bold{v}(\bold{r},t)=0$. The case of a moving solvent is discussed in the main text. In this mixture, there can be diffusive fluxes of dissolved particles. The total particle fluxes are determined as follows:
\begin{equation}
\label{macro_flux}
\bold{J}_a(\bold{r},t) =- k_{B}TL_{ab}(\{c(\bold{r},t)\}) \bold{F}_{b}(\bold{r},t)+\bold{j}_{a}(\bold{r},t),\qquad \bold{F}_a(\bold{r},t)=\nabla\left(\frac{{\mu}_a(\bold{r},t)}{k_{B}T}\right),
\end{equation}
where $L_{ab}$ is the so-called mobility matrix~\cite{akaberian2023nonequilibrium} which is symmetric, positive-definite and depends on the local concentrations of solutes, $c_a$; $\mu_a$ is the nonequilibrium intrinsic chemical potential of solute $a$; $k_B$ is the Boltzmann constant, $T$ is the temperature. In the equilibrium, where $\mu_a=const$, $\bold{F}_a=0$. As already mentioned in the main text, chemical potentials do not have a rigorous definition in nonequilibrium states. However, for systems close to equilibrium, the concept can be extended by considering the chemical potential as a functional of nonequilibrium concentrations. This method mirrors classical density functional theory (DFT), which describes equilibrium intrinsic chemical potentials as the functionals of local concentrations. Thus, we assume that the nonequilibrim chemical potentials, $\mu_a$, depend on the local concentrations as the equilibrium ones within the local equilibrium principle~\cite{kondepudi2014modern}, i.e. $\mu_{a}(\bold{r},t)=\mu_{a}(\{c_l(\bold{r},t)\})$.  Note that these chemical potentials $\mu_{a}(\bold{r},t)=\mu_{a}(\{c_l(\bold{r},t)\})$, 
of solution components depend on their local concentrations not algebraically (as in the bulk phase), but functionally -- in accordance to the classical DFT. The first term on the right-hand side of equation (\ref{macro_flux}) represents the macroscopic flux of solute $a$ in the context of nonequilibrium thermodynamics~\cite{kondepudi2014modern}. This term is attributed to the gradients of chemical potentials and is referred to as the regular part of the flux. The second term describes the random flux, which is not related to the gradients of chemical potential but arises from the thermal motion of the solute molecules -- stochastic part of the flux. This term is similar to the stochastic heat flux and stress tensor in the Landau-Lifshitz hydrodynamic fluctuation theory~\cite{landau1992hydrodynamic}.

Within each rather small volume in the vicinity of point $\bold{r}$, the total fluxes (\ref{macro_flux}) experience random fluctuations which originate from the random fluctuations of concentrations. Thus, we can write the following expression for total fluctuation of flux in the linear approximation
\begin{equation}
\delta \bold{J}_a(\bold{r},t)=\bold{J}_a(\bold{r},t)-\left<\bold{J}_a(\bold{r},t)\right>=-k_{B}T\bar{L}_{ab}\delta \bold{F}_b(\bold{r},t)+\delta\bold{j}_a(\bold{r},t),
\end{equation}
where $\bar{L}_{ab}={L}_{ab}(\{\bar{c}\})$, and $\delta \bold{F}_a=\nabla \delta\mu_a/k_BT$ is the fluctuation of the thermodynamic force, and $\delta\bold{j}_a(\bold{r},t)=\bold{j}_a(\bold{r},t)-\left<\bold{j}_a(\bold{r},t)\right>$ is the random flux. Thus the delta symbol for $\delta\bold{j}_a(\bold{r},t)$ is assigned for uniformity of notation. Note also that we have ignored the fluctuations of the mobility matrix, as these effects are of quadratic order in the concentration fluctuations. The entropy production rate~\cite{kondepudi2014modern,landau1987fluid} resulting from the relaxation of a nonequilibrium state occurring due to fluctuations of particle fluxes is
\begin{equation}
\label{rate}
\partial_t {S} = -k_{B} \int d^{3}\bold{r}\, \delta \bold{J}_a \cdot\delta \bold{F}_a,
\end{equation}
where $\partial_t =\partial/ \partial t$ and the integration is performed over the entire solution volume. Note that at equilibrium $\partial_t {S} =0$. Indeed, taking into account that $\delta \bold{F}_a = \nabla \delta \mu_a /k_B T$ and integrating by parts in (\ref{rate}), we obtain $\partial_t S=\int d^3\bold{r} \,\delta\mu_a \nabla\cdot \delta\bold{J}_a /T=0$, where we assumed that at the solution volume boundaries $\delta \bold{J}_a=0$ \footnote{Only in the case where we can ignore the loss of particles due to vaporization at the free boundary.} and that in the equilibrium state the continuous equations $\nabla\cdot \delta\bold{J}_a=0$ are fulfilled.

Thus, in present case we should understand the index $\alpha$ in general formulation (\ref{rate_production}) as multiindex $\{a,i,\bold{r}\}$ which includes the index $a$ for numbered molecular species, the index $i$ for numbered flux component, and the continuous "index" $\bold{r}$ for "numbered" small volumes in space. Thus, $\gamma_{\alpha\lambda} \to k_{B}T \bar{L}_{ab}\delta_{ik}\delta(\bold{r}-\bold{r}')$. Therefore, using basic eq. (\ref{fluctuation_dissipation}), we can obtain the correlation function of the components of the random fluxes
\begin{equation}
\label{correlator}
\left<\delta j_{a i}(\bold{r},t)\delta j_{b k}(\bold{r}',t')\right>=2\bar{L}_{ab} k_{B}T\delta_{ik}\delta(\bold{r}-\bold{r}')\delta(t-t').
\end{equation}

Note that for molecular species with spatial anisotropy, it is necessary to introduce a mobility matrix, $(L_{ik})_{ab}$ with tensor components. In general case, the fluctuation-dissipation relation for the components of the random fluxes is
\begin{equation}
\label{correlator}
\left<\delta j_{a i}(\bold{r},t)\delta j_{b k}(\bold{r}',t')\right>=2(\bar{L}_{ik})_{ab} k_{B}T\delta(\bold{r}-\bold{r}')\delta(t-t').
\end{equation}
In this paper, we only consider the isotropic case, where $(\bar{L}_{ik})_{ab} = \bar{L}_{ab} \delta_{ik}$. Note that accounting for the hydrodynamic interactions can lead to an anisotropic mobility tensor~\cite{rotne1969variational}. Note that the random fluxes have zero expectation values, i.e. $\left<\delta \bold{j}_a(\bold{r},t)\right>=0$.

\section{Equations of motion and Zwanzig equation for the bulk solutions}\label{Equations_of_motion_Zwanzig_equation_app}

Now, we can write the equations of motion for relaxation of the concentration fluctuations, $\delta c_a=c_a-\bar{c}_a$, as continuous equations
\begin{equation}
\label{discontin}
\partial_t\delta {c_a} = -\nabla \cdot\delta\bold{J}_a,
\end{equation}
which result in the following Langevin-type~\cite{brilliantov1996molekulyarnaya,zwanzig2001nonequilibrium} stochastic equations
\begin{equation}
\partial_t\delta {c_a}(\bold{r},t) = k_{B}T \bar{L}_{ab} \nabla^2 \beta \delta\mu_b(\bold{r},t) +\delta \eta_a(\bold{r},t),
\end{equation}
where $\beta=(k_BT)^{-1}$ and, as it follows from eq. (\ref{correlator}), the random noises, $\delta \eta_a=-\nabla\cdot \delta \bold{j}_a$, satisfy the following fluctuation-dissipation relation
\begin{equation}
\label{correlator_2}
\left<\delta \eta_{a}(\bold{r},t)\delta \eta_{b }(\bold{r}',t')\right>=-2\bar{L}_{ab} k_B T\nabla^2\delta(\bold{r}-\bold{r}')\delta(t-t').
\end{equation}
Taking into account that in linear approximation
\begin{equation}
\label{delta mu}
\beta\delta\mu_a(\bold{r},t) =\int d^{3}\bold{r}'\, \frac{\delta{\beta\mu}_a(\bold{r},t)}{\delta c_b(\bold{r}',t)}\bigg{|}_{c_l=\bar{c}_l}\delta c_{b}(\bold{r}',t)=\int d^{3}\bold{r}'\, C_{ab}(\bold{r}-\bold{r}')\delta c_{b}(\bold{r}',t),
\end{equation}
where we have introduced the following kernels
\begin{equation}
\label{Cab_}
C_{ab}(\bold{r},\bold{r}')=\frac{\delta{\beta\mu}_a(\bold{r},t)}{\delta c_b(\bold{r}',t)}\bigg{|}_{c_l=\bar{c}_l}=\frac{\delta^2 (\beta F)}{\delta c_a(\bold{r},t)\delta c_b(\bold{r}',t)}\bigg{|}_{c_l=\bar{c}_l}=\frac{1}{\bar{c}_a}\delta_{ab}\delta(\bold{r}-\bold{r}')- c_{ab}(\bold{r}-\bold{r}')
\end{equation}
with the standard direct correlation functions \cite{hansen2013theory,barrat2003basic,budkov2024statistical} $c_{ab}(\bold{r}-\bold{r}')$ of the species in the bulk phase,
we obtain the linear equations of motion
\begin{equation}
\label{stoch_dft}
\partial_t\delta {c}_a=-\hat{\mathcal{L}}_{ab}\delta c_{b} +\delta \eta_a,
\end{equation}
with the Liouvillian which acts in accordance with the following rule
\begin{equation}
\label{operator}
\hat{\mathcal{L}}_{ab}f_{b}(\bold{r})=- k_{B}T\bar{L}_{ac}\int d^{3}\bold{r}' \,\nabla^2C_{cb}(\bold{r},\bold{r}')f_{b}(\bold{r}').
\end{equation}
Note that in (\ref{Cab_}) we have assumed that the free energy of fluid for the equilibrium state is known,
\begin{equation}
F[\{c_a\}]=F_{id}[\{c_a\}]+F_{ex}[\{c_a\}],
\end{equation}
as the functional of the concentrations of the solutes,
where
\begin{equation}
F_{id}[\{c_a\}]=k_B T\sum\limits_{a}\int d^{3}\bold{r}\, c_{a}\left(\ln(c_a\lambda_a^3/z_a)-1\right)
\end{equation}
denotes the free energy of the ideal solution with the thermal wavelength, $\lambda_a$, of the $a$-th solute, $z_a$ is the internal partition function of the $a$-th solute molecule, and $F_{ex}$ is the excess free energy which is determined by the nature of the solution. The direct correlation functions are determined by
\begin{equation}
c_{ab}(\bold{r}-\bold{r}')=-\frac{\delta^2 (\beta F_{ex})}{\delta c_a(\bold{r})\delta c_b(\bold{r}')}\bigg{|}_{c_l=\bar{c}_l}.
\end{equation}

Equations (\ref{stoch_dft}) together with the relation (\ref{correlator_2}) can be considered as fundamental equations in the theory of relaxation kinetics of nonequilibrium concentration fluctuations of solutes in multicomponent solutions towards equilibrium.

Specifically, for a solution containing only one solute, eq. (\ref{stoch_dft}) reduces to
\begin{equation}
\label{stoch_dft_}
\partial_t\delta {c}(\bold{r},t)=-\hat{\mathcal{L}}\delta c(\bold{r},t) +\delta \eta(\bold{r},t)
\end{equation}
or in the Fourier-representation
\begin{equation}
\label{stoch_dft_+}
\partial_t\delta {c}(\bold{k},t)=-k^2\mathcal{D}(\bold{k})\delta {c}(\bold{k},t) +\delta \eta(\bold{k},t),
\end{equation}
where $k=|\bold{k}|$ and we have introduced the Fourier-images of the functions
\begin{equation}
f(\bold{k})=\int d^{3}\bold{r} \,e^{-i\bold{k}\bold{r}}f(\bold{r}),
\end{equation}
and introduced the diffusion operator in the Fourier-representation
\begin{equation}
\label{diffusion}
\mathcal{D}(\bold{k})=\bar{L} k_{B}T C(\bold{k}).
\end{equation}
Eq. (\ref{diffusion}) describes the spatial dispersion of the diffusion due to the solute-solute correlations at microscopic length scale. The diffusion coefficient is 
\begin{equation}
D=\lim\limits_{\bold{k}\to 0}\mathcal{D}(\bold{k})=\bar{L}\frac{\partial \mu}{\partial \bar{c}}
\end{equation}
that is in accordance with the macroscopic fluid mechanics definition~\cite{landau1987fluid}. We took into account that~\cite{barrat2003basic} $k_{B}T(1-\bar{c}c(0))/\bar{c}={\partial \mu}/{\partial \bar{c}}$, where $\mu$ is the chemical potential of solute. For rather dilute solutions, $\bar{L}\simeq m\bar{c}$, where $m$ is the mobility of the solute molecules, and $\mu\simeq k_{B}T\ln(\bar{c}\lambda^3/z)$, we obtain the well-known Stokes-Einstein relation, $D\simeq mk_{B}T$. It is important to note that the condition of a positive diffusion coefficient in a solution ensures the thermodynamic stability condition, i.e. the partial derivative of the chemical potential with respect to concentration is positive (${\partial \mu}/{\partial \bar{c}}>0$) (see, for instance,~\cite{landau2013statistical}). Moreover, the positive $\mathcal{D}(\mathbf{k})$ ensures the thermodynamic stability of the solution with respect to order-disorder transition (a more detailed discussion of this subject can be found in Appendix \ref{phase_stability_app}). 

For further purposes, we will require the correlation functions (correlators) of the instantaneous concentration fluctuations, 
\begin{equation}
\label{Sab_}
S_{ab}(\bold{r},\bold{r}',t)=\left<\delta c_a(\bold{r},t)\delta c_b(\bold{r}',t)\right>,
\end{equation}
which, as it follows from eq. (\ref{stoch_dft}), at $t\to\infty$ should tend to equilibrium functions $S_{ab}(\bold{r},\bold{r}')$ satisfying the following equation (see its general derivation in Zwanzig's monograph ~\cite{zwanzig2001nonequilibrium})
\begin{equation}
\label{zwanzig_1}
\hat{\mathcal{L}}_{ac} S_{cb}(\bold{r},\bold{r}')+\hat{\mathcal{L}}^{\prime}_{bc} S_{ca}(\bold{r}',\bold{r}) = -2k_{B}T \bar{L}_{ab}\nabla^2\delta(\bold{r}-\bold{r}'),
\end{equation}
where the prime above the Liouvillian indicates that it acts on the primed coordinates. We refer to this as the Zwanzig equation. Introducing the kernel of Liouvillian, $\mathcal{L}_{ab}(\bold{r},\bold{r}')=\hat{\mathcal{L}}_{ab}\delta(\bold{r}-\bold{r}')$, we can rewrite (\ref{zwanzig_1}) as
\begin{equation}
\label{zwanzig_22}
\int d^3\bold{r}^{\prime\prime}\,{\mathcal{L}}_{ac}(\bold{r},\bold{r}^{\prime\prime}) S_{cb}(\bold{r}^{\prime\prime},\bold{r}')+\int d^3\bold{r}^{\prime\prime}\,{\mathcal{L}}_{bc}(\bold{r}',\bold{r}^{\prime\prime}) S_{ca}(\bold{r}^{\prime\prime},\bold{r}) = -2 k_{B}T\bar{L}_{ab}\nabla^2\delta(\bold{r}-\bold{r}'),
\end{equation}
or in the Fourier-representation
\begin{multline}
\label{zwanzig_3}
\int \frac{d^3\bold{p}}{(2\pi)^3}{\mathcal{L}}_{ac}(\bold{k},-\bold{p}) S_{cb}(\bold{p},\bold{k}')\\+\int \frac{d^3\bold{p}}{(2\pi)^3}{\mathcal{L}}_{bc}(\bold{k}',-\bold{p}) S_{ca}(\bold{p},\bold{k}) = 2k_{B}T\bar{L}_{ab}k^2(2\pi)^3\delta(\bold{k}+\bold{k}'),
\end{multline}
where $k=|\bold{k}|$ and we have introduced the Fourier-images of the two-point functions as follows
\begin{equation}
f(\bold{k},\bold{k}')=\int d^{3}\bold{r}\int d^{3}\bold{r}' \,e^{-i\bold{k}\bold{r}-i\bold{k}'\bold{r}'}f(\bold{r},\bold{r}').
\end{equation}
In equilibrium, the bulk-phase correlation functions and Liouvillian kernels are translationally invariant, i.e. $S_{ab}(\bold{r},\bold{r}')=S_{ab}(\bold{r}-\bold{r}')$, $\mathcal{L}_{ab}(\bold{r},\bold{r}')=\mathcal{L}_{ab}(\bold{r}-\bold{r}')=- k_{B}T\bar{L}_{ac}\nabla^2C_{cb}(\bold{r}-\bold{r}')$. Consequently, their Fourier transforms become
\begin{equation}
S_{ab}(\bold{k},\bold{k}')= S_{ab}(\bold{k})(2\pi)^3 \delta(\bold{k}+\bold{k}'),
\end{equation}
\begin{equation}
\mathcal{L}_{ab}(\bold{k},\bold{k}')= k_{B}T k^2 \bar{L}_{ac}C_{cb}(\bold{k})(2\pi)^3\delta(\bold{k}+\bold{k}'),
\end{equation}
and we can rewrite (\ref{zwanzig_22}) in the Fourier-representation as
\begin{equation}
\label{zwanzig_fourier}
\bar{L}_{ac}C_{cl}(\bold{k})S_{lb}(\bold{k})+\bar{L}_{bc}C_{cl}(-\bold{k})S_{la}(-\bold{k})= 2\bar{L}_{ab}.
\end{equation}
Taking into account that the matrices $\bold{S}(\bold{k})$ and $\bold{C}(\bold{k})$ are self-conjugated, i.e. that $S_{ab}^{*}(\bold{k})=S_{ab}(-\bold{k})=S_{ba}(\bold{k})$ and $C_{ab}^{*}(\bold{k})=C_{ab}(-\bold{k})=C_{ba}(\bold{k})$, we obtain the following solution to the equation (\ref{zwanzig_fourier}):
\begin{equation}
\label{solution}
\bold{S}(\bold{k})=\bold{C}^{-1}(\bold{k}),
\end{equation}
where the matrix $\bold{C}(\bold{k})$ has the following matrix elements
\begin{equation}
C_{ab}(\bold{k})=\frac{1}{\bar{c}_a}\delta_{ab}-c_{ab}(\bold{k}).
\end{equation}
Note that self-conjugacy of matrices $\bold{S}(\bold{k})$ and $\bold{C}(\bold{k})$ follows from the following properties
\begin{equation}
S_{ab}(\bold{r},\bold{r}')=S_{ba}(\bold{r}',\bold{r}),~C_{ab}(\bold{r},\bold{r}')=C_{ba}(\bold{r}',\bold{r}),
\end{equation}
which in turn stem from the definitions (\ref{Cab_}) and (\ref{Sab_}), respectively.

Using the relation for the structure factors~\cite{hansen2013theory}  
\begin{equation}
S_{ab}(\bold{k})=\left<\delta c_a(\bold{k})\delta c_{b}(-\bold{k})\right>=\bar{c}_a\delta_{ab}+\bar{c}_a \bar{c}_b h_{ab}(\bold{k}),
\end{equation}
where $\bold{h}(\bold{k})$ is the matrix of Fourier-images of the total correlation functions, we arrive at the standard Ornstein-Zernike relation~\cite{hansen2013theory,barrat2003basic}
\begin{equation}
\label{OZ}
h_{ab}(\bold{k})=c_{ab}(\bold{k}) + c_{al}(\bold{k})X_{ls}h_{sb}(\bold{k}),\qquad X_{ls}=\bar{c}_l\delta_{ls}.
\end{equation}

\section{Phase stability of solutions}\label{phase_stability_app}

In this appendix, we briefly depart from the paper's main focus to explore a "diffusion" perspective on the theory of phase stability in solutions. Although this material extends slightly beyond the core discussion, it offers valuable insights into the interplay between the kinetic theory of diffusion transport in solutions -- rooted in the principles of nonequilibrium statistical physics -- and the conventional understanding of phase stability typically associated with equilibrium theory. This exploration will enhance our comprehension of these interconnected concepts~\cite{landau1937towards,kats1993weak,borue1988statistical,de1991transitions,khokhlov1992compatibility}.

Let us average eq. (\ref{stoch_dft}), taking into account that $\left<\delta\eta_a(\bold{r},t)\right>=0$, and write it in the Fourier representation as follows 
\begin{equation}
\label{average_stoch_dft}
\partial_t\left<\delta c_{a}(\bold{k},t)\right>=-\mathcal{L}_{ab}(\bold{k})\left<\delta c_{b}(\bold{k},t)\right>, 
\end{equation}
where $\mathcal{L}_{ab}(\bold{k})=k_{B}T k^2 \bar{L}_{ac} C_{cb}(\bold{k})$.
To solve equation (\ref{average_stoch_dft}), we have to determine the initial conditions $\left<\delta c_{a}(\bold{k},0)\right>=g_{a}(\bold{k})$, which describe the initial disturbances in the concentrations of the solutes. The solution is
\begin{equation}
\left<\delta c_a(\bold{k},t)\right>=U_{ab}(\bold{k},t) g_b(\bold{k}),
\end{equation}
where we have introduced the following "evolution" matrix
\begin{equation}
\bold{U}(\bold{k},t)=e^{-\mathcal{L}(\bold{k})t}.
\end{equation}
If the phase of solution is stable then any small changes in the concentrations will cause them to return to their original average values at $t\to \infty$. In order for the average fluctuations in the concentrations of the species tend to zero, it is necessary for the matrix $\mathcal{L}_{ab}(\bold{k})$ to be positive-definite, i.e. all its eigenvalues $\lambda_{a}(\bold{k})>0$. In particular, it leads to the necessary condition $\det\mathcal{L}(\bold{k})>0$, which can be rewritten as the standard known condition of the phase stability~\cite{khokhlov1992compatibility}
\begin{equation}
\label{stability}
\det\bold{S}^{-1}(\bold{k})>0,
\end{equation}
where we took into account eq. (\ref{solution}).

If for a given set of the system's parameters the inequality (\ref{stability}) is violated for some values of the wave vector
$\bold{k}$, the system becomes unstable and the spinodal decomposition begins~\cite{archer2004dynamical}. If inequality (\ref{stability}) is first violated at $\bold{k}=0$, we obtain the instability with respect to spinodal decomposition into macrophases. In particular, in the case when only one solute is dissolved, eq. (\ref{stability}) reduces to $S^{-1}(\bold{k})>0$ which for $\bold{k}=0$ leads to already mentioned in the main text condition of solution thermodynamic stability, $\partial \mu /\partial\bar{c} >0$. This is related to the fact that $S^{-1}(0)=\partial{(\beta \mu)}/\partial{\bar{c}}$. In case of more than one dissolved species the thermodynamic stability condition implies positive-definite matrix with elements $S^{-1}_{ab}(0)=\partial{(\beta \mu_a)}/\partial{\bar{c}_b}$. If the inequality (\ref{stability}) holds at $\bold{k}=0$ but not for certain wave-vector $\bold{k}=\bold{k}_{0}\neq 0$, spinodal decomposition occurs with the formation of microscopic domain structures with a characteristic spatial size of $d_0\simeq 2\pi/|\bold{k}_{0}|$. Such emerging superstructures can be observed in various condensed matter systems, including polymer solutions and melts~\cite{borue1988statistical,de1991transitions,khokhlov1992compatibility}, electrolyte solutions~\cite{nabutovskii1980order,hoye1990effect}, alloys~\cite{cook1970brownian,langer1971theory,zakharov1978impurity}, among others~\cite{skripov1979spinodal}. This phenomenon is called the order-disorder transition, or weak crystallization~\cite{landau1937towards,brazovskii1987theory,kats1993weak,erukhimovich2019thermodynamics}.

This statement can be justified from another perspective. Indeed, when $t\to \infty$, the kinetic equations (\ref{average_stoch_dft}), describing average concentration fluctuations, transform into a homogeneous system of linear equations
\begin{equation}
\label{hom_syst}
\mathcal{L}_{ab}(\bold{k})\left<\delta c_{b}(\bold{k})\right>=0.
\end{equation}
If the condition (\ref{stability}) is fulfilled, which in turn implies that $\det\mathcal{L}(\bold{k})\neq 0$, the system (\ref{hom_syst}) has only trivial solution $\left<\delta c_{a}(\bold{k})\right>=0$. However, when $\det\mathcal{L}(\bold{k})=\det\bold{\bar{L}}\det\bold{S}^{-1}(\bold{k})=0$, as is well known from the linear algebra, solutions $\left<\delta c_{a}(\bold{k})\right>\neq 0$ are possible. Therefore, solution $\bold{k}_{0}$ to equation $\det \bold{S}^{-1}(\bold{k})=0$ is the wave-vector describing the emerging ordered structure, occurring in a certain solution due to the spinodal decomposition. Thus, the solutions of (\ref{hom_syst})
\begin{equation}
\label{solution_2}
\left<\delta c_{a}(\bold{k})\right>=A_a(\bold{k}_0)\delta(\bold{k}-\bold{k}_{0}),
\end{equation}
where $A_a= C_a e^{i\psi_a}$ is a complex constant that depends on the "critical" wave vector $\bold{k}_0$.

The solution (\ref{solution_2}) in coordinate space is
\begin{equation}
\left<\delta c_{a}(\bold{r})\right>=\sum\limits_{\bold{k}\in \mathcal{R}^{-1}}\text{Re}\left(A_a e^{i\bold{k}\bold{r}}\right)=\sum\limits_{\bold{k}\in \mathcal{R}^{-1}}C_a \cos(\bold{k}\bold{r}+\psi_a),
\end{equation}
where the summation is taken over the nodes that belong to the reciprocal lattice, $\mathcal{R}^{-1}$, of the chosen Bravais lattice, $\mathcal{R}$, with the condition $|\mathbf{k}| = k_0$ being obeyed. The preferred symmetry of the ordered phase of the Bravais lattice can be determined by minimizing the free energy using theoretical background developed in papers~\cite{brazovskii1987theory,kats1993weak,de1991transitions,erukhimovich2019thermodynamics}.

\section{Derivation of eq. (\ref{delta_eta})}\label{eq_delta_eta_app}

In this appendix, we derive the expression for the contribution of solute-solute interactions to the solution viscosity, which is given by equation (\ref{delta_eta}) in the main text. We start with the Fourier transform of the average internal force density 
\begin{equation}
\label{bar_G}
\bar{\bold{G}}_{\text{int}}(\bold{k})=\frac{1}{\bar{L}}\int\frac{d^3\bold{q}}{(2\pi)^3}\frac{(S(\bold{q})-S(\bold{k}-\bold{q}))S(\bold{q})S(\bold{k}-\bold{q})}{(\bold{k}-\bold{q})^2S(\bold{q})+\bold{q}^2S(\bold{k}-\bold{q})}c_2(\bold{q})(\bold{q}\otimes\bold{q}):\bar{\bold{v}}(\bold{k}).
\end{equation}
Assuming that the structure factor depends only on modulus of $\bold{k}$, i.e. that 
\begin{equation}
S(\bold{q})=s(q^2),~S(\bold{k}-\bold{q})=s(q^2+k^2-2\bold{k}\cdot\bold{q}),
\end{equation}
we obtain up to quadratic terms in $k$
\begin{multline}
\frac{(S(\bold{q})-S(\bold{k}-\bold{q}))S(\bold{q})S(\bold{k}-\bold{q})}{(\bold{k}-\bold{q})^2S(\bold{q})+\bold{q}^2S(\bold{k}-\bold{q})}\\\sim -\frac{s(q^2)}{2q^2}\left(k^2s'(q^2)+2s''(q^2)(\bold{k}\cdot \bold{q})^2+\frac{2s'(q^2)^2}{s(q^2)}(\bold{k}\cdot \bold{q})^2-2s'(q^2)\frac{(\bold{k}\cdot \bold{q})^2}{q^2}\right),
\end{multline}
where we have omitted the linear term in $\bold{k}$ since they drop out after integration over $d^3\bold{q}$ in (\ref{bar_G}). Considering that $\bold{k} \cdot \bar{\bold{v}}(\bold{k}) = 0$, we can assume that the vector $\bold{k}$ is aligned with the $z$-axis, and the vector $\bar{\bold{v}}(\bold{k})$ is aligned with the $x$-axis in $xy$-plain which is orthogonal to $z$-axis. Therefore, using the spherical coordinates $(q,\vartheta,\phi)$ in inverse space, we obtain at small $k$
\begin{multline}
\bar{\bold{G}}_{\text{int}}(\bold{k})\simeq -\frac{k^2 \bar{\bold{v}}(\bold{k})}{2\bar{L}}\int\limits_{0}^{\infty}\frac{dq\,q^2}{8\pi^3} \int\limits_{0}^{\pi}d\vartheta \sin\vartheta\int\limits_{0}^{2\pi}d\phi\cos^2\phi\sin^2\vartheta \\\times c_2(q)s(q^2)\left(s'(q^2)(1-2\cos^2\vartheta)+2s''(q^2)q^2\cos^2\vartheta+\frac{2s'(q^2)^2}{s(q^2)}q^2\cos^2\vartheta\right)\\=-\frac{k^2 \bar{\bold{v}}(\bold{k})}{30\pi^2 \bar{L}}\int\limits_{0}^{\infty}dq \,q^2
c_2(q)s(q^2)\left(\frac{3}{2}s'(q^2)+q^2s''(q^2)+\frac{q^2s'(q^2)^2}{s(q^2)}\right).
\end{multline}

Therefore, we obtain the following expression for the viscosity
\begin{equation}
\label{delta_eta_intermed}
\Delta\eta =\frac{1}{30\pi^2 \bar{L}}\int\limits_{0}^{\infty}dq\,q^2
c_2(q)s(q^2)\left(\frac{3}{2}s'(q^2)+q^2s''(q^2)+\frac{q^2s'(q^2)^2}{s(q^2)}\right),
\end{equation}
where we have used that $c_2(\bold{q})=c_2(q)$. Eq. (\ref{delta_eta_intermed}) can be simplified further. Indeed, by accounting for $c_2(q)s(q^2)=\bar{c}h_2(q)$, $s(q^2)=\bar{c}(1+\bar{c}h_2(q))$, $s'(q^2)=\bar{c}^2h_2'(q)/2 q$, $s''(q^2)=\bar{c}^2\left(h_2''(q)-h_2'(q)/q\right)/4q^2$, we obtain
\begin{equation}
\label{delta_eta_intermed_}
\Delta\eta =\frac{\bar{c}^3}{30\pi^2 \bar{L}}\int\limits_{0}^{\infty}dq\,q^2
h_2(q)\left(\frac{h_2'(q)}{2q}+\frac{h_2''(q)}{4}+\frac{\bar{c}h_2'(q)^2}{4(1+\bar{c}h_2(q))}\right).
\end{equation}

Finally, after integrating by parts the second term in the integrand of eq. (\ref{delta_eta_intermed_}) after some algebra, we arrive at eq. (\ref{delta_eta}).

\selectlanguage{english}
\bibliography{name}

\begin{thebibliography}{78}%
\makeatletter
\providecommand \@ifxundefined [1]{%
 \@ifx{#1\undefined}
}%
\providecommand \@ifnum [1]{%
 \ifnum #1\expandafter \@firstoftwo
 \else \expandafter \@secondoftwo
 \fi
}%
\providecommand \@ifx [1]{%
 \ifx #1\expandafter \@firstoftwo
 \else \expandafter \@secondoftwo
 \fi
}%
\providecommand \natexlab [1]{#1}%
\providecommand \enquote  [1]{``#1''}%
\providecommand \bibnamefont  [1]{#1}%
\providecommand \bibfnamefont [1]{#1}%
\providecommand \citenamefont [1]{#1}%
\providecommand \href@noop [0]{\@secondoftwo}%
\providecommand \href [0]{\begingroup \@sanitize@url \@href}%
\providecommand \@href[1]{\@@startlink{#1}\@@href}%
\providecommand \@@href[1]{\endgroup#1\@@endlink}%
\providecommand \@sanitize@url [0]{\catcode `\\12\catcode `\$12\catcode `\&12\catcode `\#12\catcode `\^12\catcode `\_12\catcode `\%12\relax}%
\providecommand \@@startlink[1]{}%
\providecommand \@@endlink[0]{}%
\providecommand \url  [0]{\begingroup\@sanitize@url \@url }%
\providecommand \@url [1]{\endgroup\@href {#1}{\urlprefix }}%
\providecommand \urlprefix  [0]{URL }%
\providecommand \Eprint [0]{\href }%
\providecommand \doibase [0]{https://doi.org/}%
\providecommand \selectlanguage [0]{\@gobble}%
\providecommand \bibinfo  [0]{\@secondoftwo}%
\providecommand \bibfield  [0]{\@secondoftwo}%
\providecommand \translation [1]{[#1]}%
\providecommand \BibitemOpen [0]{}%
\providecommand \bibitemStop [0]{}%
\providecommand \bibitemNoStop [0]{.\EOS\space}%
\providecommand \EOS [0]{\spacefactor3000\relax}%
\providecommand \BibitemShut  [1]{\csname bibitem#1\endcsname}%
\let\auto@bib@innerbib\@empty
\bibitem [{\citenamefont {de~Las~Heras}\ \emph {et~al.}(2023)\citenamefont {de~Las~Heras}, \citenamefont {Zimmermann}, \citenamefont {Samm{\"u}ller}, \citenamefont {Hermann},\ and\ \citenamefont {Schmidt}}]{de2023perspective}%
  \BibitemOpen
  \bibfield  {author} {\bibinfo {author} {\bibfnamefont {D.}~\bibnamefont {de~Las~Heras}}, \bibinfo {author} {\bibfnamefont {T.}~\bibnamefont {Zimmermann}}, \bibinfo {author} {\bibfnamefont {F.}~\bibnamefont {Samm{\"u}ller}}, \bibinfo {author} {\bibfnamefont {S.}~\bibnamefont {Hermann}},\ and\ \bibinfo {author} {\bibfnamefont {M.}~\bibnamefont {Schmidt}},\ }\bibfield  {title} {\enquote {\bibinfo {title} {Perspective: How to overcome dynamical density functional theory},}\ }\href@noop {} {\bibfield  {journal} {\bibinfo  {journal} {Journal of Physics: Condensed Matter}\ }\textbf {\bibinfo {volume} {35}},\ \bibinfo {pages} {271501} (\bibinfo {year} {2023})}\BibitemShut {NoStop}%
\bibitem [{\citenamefont {Mills-Williams}, \citenamefont {Goddard},\ and\ \citenamefont {Archer}(2024)}]{mills2024dynamic}%
  \BibitemOpen
  \bibfield  {author} {\bibinfo {author} {\bibfnamefont {R.~D.}\ \bibnamefont {Mills-Williams}}, \bibinfo {author} {\bibfnamefont {B.~D.}\ \bibnamefont {Goddard}},\ and\ \bibinfo {author} {\bibfnamefont {A.~J.}\ \bibnamefont {Archer}},\ }\bibfield  {title} {\enquote {\bibinfo {title} {Dynamic density functional theory with inertia and background flow},}\ }\href@noop {} {\bibfield  {journal} {\bibinfo  {journal} {The Journal of Chemical Physics}\ }\textbf {\bibinfo {volume} {160}} (\bibinfo {year} {2024})}\BibitemShut {NoStop}%
\bibitem [{\citenamefont {Illien}(2024)}]{illien2024dean}%
  \BibitemOpen
  \bibfield  {author} {\bibinfo {author} {\bibfnamefont {P.}~\bibnamefont {Illien}},\ }\bibfield  {title} {\enquote {\bibinfo {title} {The dean-kawasaki equation and stochastic density functional theory},}\ }\href@noop {} {\bibfield  {journal} {\bibinfo  {journal} {arXiv preprint arXiv:2411.13467}\ } (\bibinfo {year} {2024})}\BibitemShut {NoStop}%
\bibitem [{\citenamefont {Hansen}\ and\ \citenamefont {McDonald}(2013)}]{hansen2013theory}%
  \BibitemOpen
  \bibfield  {author} {\bibinfo {author} {\bibfnamefont {J.-P.}\ \bibnamefont {Hansen}}\ and\ \bibinfo {author} {\bibfnamefont {I.~R.}\ \bibnamefont {McDonald}},\ }\href@noop {} {\emph {\bibinfo {title} {Theory of simple liquids: with applications to soft matter}}}\ (\bibinfo  {publisher} {Academic press},\ \bibinfo {year} {2013})\BibitemShut {NoStop}%
\bibitem [{\citenamefont {Barrat}\ and\ \citenamefont {Hansen}(2003)}]{barrat2003basic}%
  \BibitemOpen
  \bibfield  {author} {\bibinfo {author} {\bibfnamefont {J.-L.}\ \bibnamefont {Barrat}}\ and\ \bibinfo {author} {\bibfnamefont {J.-P.}\ \bibnamefont {Hansen}},\ }\href@noop {} {\emph {\bibinfo {title} {Basic concepts for simple and complex liquids}}}\ (\bibinfo  {publisher} {Cambridge University Press},\ \bibinfo {year} {2003})\BibitemShut {NoStop}%
\bibitem [{\citenamefont {Rex}\ and\ \citenamefont {L{\"o}wen}(2008)}]{rex2008dynamical}%
  \BibitemOpen
  \bibfield  {author} {\bibinfo {author} {\bibfnamefont {M.}~\bibnamefont {Rex}}\ and\ \bibinfo {author} {\bibfnamefont {H.}~\bibnamefont {L{\"o}wen}},\ }\bibfield  {title} {\enquote {\bibinfo {title} {Dynamical density functional theory with hydrodynamic interactions and colloids in unstable traps},}\ }\href@noop {} {\bibfield  {journal} {\bibinfo  {journal} {Physical review letters}\ }\textbf {\bibinfo {volume} {101}},\ \bibinfo {pages} {148302} (\bibinfo {year} {2008})}\BibitemShut {NoStop}%
\bibitem [{\citenamefont {Donev}\ and\ \citenamefont {Vanden-Eijnden}(2014)}]{donev2014dynamic}%
  \BibitemOpen
  \bibfield  {author} {\bibinfo {author} {\bibfnamefont {A.}~\bibnamefont {Donev}}\ and\ \bibinfo {author} {\bibfnamefont {E.}~\bibnamefont {Vanden-Eijnden}},\ }\bibfield  {title} {\enquote {\bibinfo {title} {Dynamic density functional theory with hydrodynamic interactions and fluctuations},}\ }\href@noop {} {\bibfield  {journal} {\bibinfo  {journal} {The Journal of chemical physics}\ }\textbf {\bibinfo {volume} {140}} (\bibinfo {year} {2014})}\BibitemShut {NoStop}%
\bibitem [{\citenamefont {Aerov}\ and\ \citenamefont {Kr{\"u}ger}(2015)}]{aerov2015theory}%
  \BibitemOpen
  \bibfield  {author} {\bibinfo {author} {\bibfnamefont {A.~A.}\ \bibnamefont {Aerov}}\ and\ \bibinfo {author} {\bibfnamefont {M.}~\bibnamefont {Kr{\"u}ger}},\ }\bibfield  {title} {\enquote {\bibinfo {title} {Theory of rheology in confinement},}\ }\href@noop {} {\bibfield  {journal} {\bibinfo  {journal} {Physical Review E}\ }\textbf {\bibinfo {volume} {92}},\ \bibinfo {pages} {042301} (\bibinfo {year} {2015})}\BibitemShut {NoStop}%
\bibitem [{\citenamefont {Dean}(1996)}]{dean1996langevin}%
  \BibitemOpen
  \bibfield  {author} {\bibinfo {author} {\bibfnamefont {D.~S.}\ \bibnamefont {Dean}},\ }\bibfield  {title} {\enquote {\bibinfo {title} {Langevin equation for the density of a system of interacting langevin processes},}\ }\href@noop {} {\bibfield  {journal} {\bibinfo  {journal} {Journal of Physics A: Mathematical and General}\ }\textbf {\bibinfo {volume} {29}},\ \bibinfo {pages} {L613} (\bibinfo {year} {1996})}\BibitemShut {NoStop}%
\bibitem [{\citenamefont {D{\'e}mery}\ and\ \citenamefont {Dean}(2016)}]{demery2016conductivity}%
  \BibitemOpen
  \bibfield  {author} {\bibinfo {author} {\bibfnamefont {V.}~\bibnamefont {D{\'e}mery}}\ and\ \bibinfo {author} {\bibfnamefont {D.~S.}\ \bibnamefont {Dean}},\ }\bibfield  {title} {\enquote {\bibinfo {title} {The conductivity of strong electrolytes from stochastic density functional theory},}\ }\href@noop {} {\bibfield  {journal} {\bibinfo  {journal} {Journal of Statistical Mechanics: Theory and Experiment}\ }\textbf {\bibinfo {volume} {2016}},\ \bibinfo {pages} {023106} (\bibinfo {year} {2016})}\BibitemShut {NoStop}%
\bibitem [{\citenamefont {Avni}\ \emph {et~al.}(2022)\citenamefont {Avni}, \citenamefont {Adar}, \citenamefont {Andelman},\ and\ \citenamefont {Orland}}]{avni2022conductivity}%
  \BibitemOpen
  \bibfield  {author} {\bibinfo {author} {\bibfnamefont {Y.}~\bibnamefont {Avni}}, \bibinfo {author} {\bibfnamefont {R.~M.}\ \bibnamefont {Adar}}, \bibinfo {author} {\bibfnamefont {D.}~\bibnamefont {Andelman}},\ and\ \bibinfo {author} {\bibfnamefont {H.}~\bibnamefont {Orland}},\ }\bibfield  {title} {\enquote {\bibinfo {title} {Conductivity of concentrated electrolytes},}\ }\href@noop {} {\bibfield  {journal} {\bibinfo  {journal} {Physical Review Letters}\ }\textbf {\bibinfo {volume} {128}},\ \bibinfo {pages} {098002} (\bibinfo {year} {2022})}\BibitemShut {NoStop}%
\bibitem [{\citenamefont {Avni}, \citenamefont {Andelman},\ and\ \citenamefont {Orland}(2022)}]{avni2022conductance}%
  \BibitemOpen
  \bibfield  {author} {\bibinfo {author} {\bibfnamefont {Y.}~\bibnamefont {Avni}}, \bibinfo {author} {\bibfnamefont {D.}~\bibnamefont {Andelman}},\ and\ \bibinfo {author} {\bibfnamefont {H.}~\bibnamefont {Orland}},\ }\bibfield  {title} {\enquote {\bibinfo {title} {Conductance of concentrated electrolytes: Multivalency and the wien effect},}\ }\href@noop {} {\bibfield  {journal} {\bibinfo  {journal} {The Journal of Chemical Physics}\ }\textbf {\bibinfo {volume} {157}} (\bibinfo {year} {2022})}\BibitemShut {NoStop}%
\bibitem [{\citenamefont {Robin}(2024)}]{robin2024correlation}%
  \BibitemOpen
  \bibfield  {author} {\bibinfo {author} {\bibfnamefont {P.}~\bibnamefont {Robin}},\ }\bibfield  {title} {\enquote {\bibinfo {title} {Correlation-induced viscous dissipation in concentrated electrolytes},}\ }\href@noop {} {\bibfield  {journal} {\bibinfo  {journal} {The Journal of Chemical Physics}\ }\textbf {\bibinfo {volume} {160}} (\bibinfo {year} {2024})}\BibitemShut {NoStop}%
\bibitem [{\citenamefont {Zwanzig}(2001)}]{zwanzig2001nonequilibrium}%
  \BibitemOpen
  \bibfield  {author} {\bibinfo {author} {\bibfnamefont {R.}~\bibnamefont {Zwanzig}},\ }\href@noop {} {\emph {\bibinfo {title} {Nonequilibrium statistical mechanics}}}\ (\bibinfo  {publisher} {Oxford university press},\ \bibinfo {year} {2001})\BibitemShut {NoStop}%
\bibitem [{\citenamefont {Akaberian}\ \emph {et~al.}(2023)\citenamefont {Akaberian}, \citenamefont {Thewes}, \citenamefont {Sollich},\ and\ \citenamefont {Kr{\"u}ger}}]{akaberian2023nonequilibrium}%
  \BibitemOpen
  \bibfield  {author} {\bibinfo {author} {\bibfnamefont {M.}~\bibnamefont {Akaberian}}, \bibinfo {author} {\bibfnamefont {F.~C.}\ \bibnamefont {Thewes}}, \bibinfo {author} {\bibfnamefont {P.}~\bibnamefont {Sollich}},\ and\ \bibinfo {author} {\bibfnamefont {M.}~\bibnamefont {Kr{\"u}ger}},\ }\bibfield  {title} {\enquote {\bibinfo {title} {Nonequilibrium mixture dynamics: A model for mobilities and its consequences},}\ }\href@noop {} {\bibfield  {journal} {\bibinfo  {journal} {The Journal of Chemical Physics}\ }\textbf {\bibinfo {volume} {158}} (\bibinfo {year} {2023})}\BibitemShut {NoStop}%
\bibitem [{\citenamefont {Archer}\ and\ \citenamefont {Evans}(2001)}]{archer2001binary}%
  \BibitemOpen
  \bibfield  {author} {\bibinfo {author} {\bibfnamefont {A.}~\bibnamefont {Archer}}\ and\ \bibinfo {author} {\bibfnamefont {R.}~\bibnamefont {Evans}},\ }\bibfield  {title} {\enquote {\bibinfo {title} {Binary gaussian core model: Fluid-fluid phase separation and interfacial properties},}\ }\href@noop {} {\bibfield  {journal} {\bibinfo  {journal} {Physical Review E}\ }\textbf {\bibinfo {volume} {64}},\ \bibinfo {pages} {041501} (\bibinfo {year} {2001})}\BibitemShut {NoStop}%
\bibitem [{\citenamefont {Budkov}\ and\ \citenamefont {Kalikin}(2024)}]{budkov2024statistical}%
  \BibitemOpen
  \bibfield  {author} {\bibinfo {author} {\bibfnamefont {Y.~A.}\ \bibnamefont {Budkov}}\ and\ \bibinfo {author} {\bibfnamefont {N.~N.}\ \bibnamefont {Kalikin}},\ }\href@noop {} {\emph {\bibinfo {title} {Statistical Field Theory of Ion-Molecular Fluids}}}\ (\bibinfo  {publisher} {Springer},\ \bibinfo {year} {2024})\BibitemShut {NoStop}%
\bibitem [{\citenamefont {Kondepudi}\ and\ \citenamefont {Prigogine}(2014)}]{kondepudi2014modern}%
  \BibitemOpen
  \bibfield  {author} {\bibinfo {author} {\bibfnamefont {D.}~\bibnamefont {Kondepudi}}\ and\ \bibinfo {author} {\bibfnamefont {I.}~\bibnamefont {Prigogine}},\ }\href@noop {} {\emph {\bibinfo {title} {Modern thermodynamics: from heat engines to dissipative structures}}}\ (\bibinfo  {publisher} {John wiley \& sons},\ \bibinfo {year} {2014})\BibitemShut {NoStop}%
\bibitem [{\citenamefont {Klimontovich}(2024)}]{klimontovich2024statistical}%
  \BibitemOpen
  \bibfield  {author} {\bibinfo {author} {\bibfnamefont {Y.}~\bibnamefont {Klimontovich}},\ }\href@noop {} {\emph {\bibinfo {title} {Statistical physics}}}\ (\bibinfo  {publisher} {CRC Press},\ \bibinfo {year} {2024})\BibitemShut {NoStop}%
\bibitem [{\citenamefont {Landau}\ and\ \citenamefont {Lifshitz}(2013)}]{landau2013statistical}%
  \BibitemOpen
  \bibfield  {author} {\bibinfo {author} {\bibfnamefont {L.~D.}\ \bibnamefont {Landau}}\ and\ \bibinfo {author} {\bibfnamefont {E.~M.}\ \bibnamefont {Lifshitz}},\ }\href@noop {} {\emph {\bibinfo {title} {Statistical Physics}}},\ Vol.~\bibinfo {volume} {5}\ (\bibinfo  {publisher} {Elsevier},\ \bibinfo {year} {2013})\BibitemShut {NoStop}%
\bibitem [{\citenamefont {Evans}(2009)}]{evans2009density}%
  \BibitemOpen
  \bibfield  {author} {\bibinfo {author} {\bibfnamefont {R.}~\bibnamefont {Evans}},\ }\bibfield  {title} {\enquote {\bibinfo {title} {Density functional theory for inhomogeneous fluids i: Simple fluids in equilibrium},}\ }\href@noop {} {\bibfield  {journal} {\bibinfo  {journal} {Lectures at 3rd Warsaw School of Statistical Physics, Kazimierz Dolny}\ }\textbf {\bibinfo {volume} {27}} (\bibinfo {year} {2009})}\BibitemShut {NoStop}%
\bibitem [{\citenamefont {Samm{\"u}ller}, \citenamefont {Schmidt},\ and\ \citenamefont {Evans}(2025)}]{sammuller2025neural}%
  \BibitemOpen
  \bibfield  {author} {\bibinfo {author} {\bibfnamefont {F.}~\bibnamefont {Samm{\"u}ller}}, \bibinfo {author} {\bibfnamefont {M.}~\bibnamefont {Schmidt}},\ and\ \bibinfo {author} {\bibfnamefont {R.}~\bibnamefont {Evans}},\ }\bibfield  {title} {\enquote {\bibinfo {title} {Neural density functional theory of liquid-gas phase coexistence},}\ }\href@noop {} {\bibfield  {journal} {\bibinfo  {journal} {Physical Review X}\ }\textbf {\bibinfo {volume} {15}},\ \bibinfo {pages} {011013} (\bibinfo {year} {2025})}\BibitemShut {NoStop}%
\bibitem [{\citenamefont {Maduar}\ \emph {et~al.}(2015)\citenamefont {Maduar}, \citenamefont {Belyaev}, \citenamefont {Lobaskin},\ and\ \citenamefont {Vinogradova}}]{maduar2015electrohydrodynamics}%
  \BibitemOpen
  \bibfield  {author} {\bibinfo {author} {\bibfnamefont {S.}~\bibnamefont {Maduar}}, \bibinfo {author} {\bibfnamefont {A.}~\bibnamefont {Belyaev}}, \bibinfo {author} {\bibfnamefont {V.}~\bibnamefont {Lobaskin}},\ and\ \bibinfo {author} {\bibfnamefont {O.}~\bibnamefont {Vinogradova}},\ }\bibfield  {title} {\enquote {\bibinfo {title} {Electrohydrodynamics near hydrophobic surfaces},}\ }\href@noop {} {\bibfield  {journal} {\bibinfo  {journal} {Physical review letters}\ }\textbf {\bibinfo {volume} {114}},\ \bibinfo {pages} {118301} (\bibinfo {year} {2015})}\BibitemShut {NoStop}%
\bibitem [{\citenamefont {Vinogradova}\ and\ \citenamefont {Silkina}(2023)}]{vinogradova2023electrophoresis}%
  \BibitemOpen
  \bibfield  {author} {\bibinfo {author} {\bibfnamefont {O.~I.}\ \bibnamefont {Vinogradova}}\ and\ \bibinfo {author} {\bibfnamefont {E.~F.}\ \bibnamefont {Silkina}},\ }\bibfield  {title} {\enquote {\bibinfo {title} {Electrophoresis of ions and electrolyte conductivity: From bulk to nanochannels},}\ }\href@noop {} {\bibfield  {journal} {\bibinfo  {journal} {The Journal of Chemical Physics}\ }\textbf {\bibinfo {volume} {159}} (\bibinfo {year} {2023})}\BibitemShut {NoStop}%
\bibitem [{\citenamefont {Golestanian}(2025)}]{golestanian2025hydrodynamically}%
  \BibitemOpen
  \bibfield  {author} {\bibinfo {author} {\bibfnamefont {R.}~\bibnamefont {Golestanian}},\ }\bibfield  {title} {\enquote {\bibinfo {title} {Hydrodynamically consistent many-body harada-sasa relation},}\ }\href@noop {} {\bibfield  {journal} {\bibinfo  {journal} {Physical Review Letters}\ }\textbf {\bibinfo {volume} {134}},\ \bibinfo {pages} {207101} (\bibinfo {year} {2025})}\BibitemShut {NoStop}%
\bibitem [{\citenamefont {De~Zarate}\ and\ \citenamefont {Sengers}(2006)}]{de2006hydrodynamic}%
  \BibitemOpen
  \bibfield  {author} {\bibinfo {author} {\bibfnamefont {J.~M.~O.}\ \bibnamefont {De~Zarate}}\ and\ \bibinfo {author} {\bibfnamefont {J.~V.}\ \bibnamefont {Sengers}},\ }\href@noop {} {\emph {\bibinfo {title} {Hydrodynamic fluctuations in fluids and fluid mixtures}}}\ (\bibinfo  {publisher} {Elsevier},\ \bibinfo {year} {2006})\BibitemShut {NoStop}%
\bibitem [{\citenamefont {Landau}\ \emph {et~al.}(1992)\citenamefont {Landau}, \citenamefont {Lifshitz}, \citenamefont {Beyer} \emph {et~al.}}]{landau1992hydrodynamic}%
  \BibitemOpen
  \bibfield  {author} {\bibinfo {author} {\bibfnamefont {L.~D.}\ \bibnamefont {Landau}}, \bibinfo {author} {\bibfnamefont {E.}~\bibnamefont {Lifshitz}}, \bibinfo {author} {\bibfnamefont {R.}~\bibnamefont {Beyer}}, \emph {et~al.},\ }\bibfield  {title} {\enquote {\bibinfo {title} {Hydrodynamic fluctuations},}\ }in\ \href@noop {} {\emph {\bibinfo {booktitle} {Perspectives in Theoretical Physics}}}\ (\bibinfo  {publisher} {Elsevier},\ \bibinfo {year} {1992})\ pp.\ \bibinfo {pages} {359--361}\BibitemShut {NoStop}%
\bibitem [{\citenamefont {Schmidt}(2022)}]{schmidt2022power}%
  \BibitemOpen
  \bibfield  {author} {\bibinfo {author} {\bibfnamefont {M.}~\bibnamefont {Schmidt}},\ }\bibfield  {title} {\enquote {\bibinfo {title} {Power functional theory for many-body dynamics},}\ }\href@noop {} {\bibfield  {journal} {\bibinfo  {journal} {Reviews of Modern Physics}\ }\textbf {\bibinfo {volume} {94}},\ \bibinfo {pages} {015007} (\bibinfo {year} {2022})}\BibitemShut {NoStop}%
\bibitem [{\citenamefont {Cui}\ and\ \citenamefont {Zaccone}(2018)}]{cui2018generalized}%
  \BibitemOpen
  \bibfield  {author} {\bibinfo {author} {\bibfnamefont {B.}~\bibnamefont {Cui}}\ and\ \bibinfo {author} {\bibfnamefont {A.}~\bibnamefont {Zaccone}},\ }\bibfield  {title} {\enquote {\bibinfo {title} {Generalized langevin equation and fluctuation-dissipation theorem for particle-bath systems in external oscillating fields},}\ }\href@noop {} {\bibfield  {journal} {\bibinfo  {journal} {Physical Review E}\ }\textbf {\bibinfo {volume} {97}},\ \bibinfo {pages} {060102} (\bibinfo {year} {2018})}\BibitemShut {NoStop}%
\bibitem [{\citenamefont {Pelargonio}\ and\ \citenamefont {Zaccone}(2023)}]{pelargonio2023generalized}%
  \BibitemOpen
  \bibfield  {author} {\bibinfo {author} {\bibfnamefont {S.}~\bibnamefont {Pelargonio}}\ and\ \bibinfo {author} {\bibfnamefont {A.}~\bibnamefont {Zaccone}},\ }\bibfield  {title} {\enquote {\bibinfo {title} {Generalized langevin equation with shear flow and its fluctuation-dissipation theorems derived from a caldeira-leggett hamiltonian},}\ }\href@noop {} {\bibfield  {journal} {\bibinfo  {journal} {Physical Review E}\ }\textbf {\bibinfo {volume} {107}},\ \bibinfo {pages} {064102} (\bibinfo {year} {2023})}\BibitemShut {NoStop}%
\bibitem [{\citenamefont {Kr{\"u}ger}\ and\ \citenamefont {Dean}(2017)}]{kruger2017gaussian}%
  \BibitemOpen
  \bibfield  {author} {\bibinfo {author} {\bibfnamefont {M.}~\bibnamefont {Kr{\"u}ger}}\ and\ \bibinfo {author} {\bibfnamefont {D.~S.}\ \bibnamefont {Dean}},\ }\bibfield  {title} {\enquote {\bibinfo {title} {A gaussian theory for fluctuations in simple liquids},}\ }\href@noop {} {\bibfield  {journal} {\bibinfo  {journal} {The Journal of Chemical Physics}\ }\textbf {\bibinfo {volume} {146}} (\bibinfo {year} {2017})}\BibitemShut {NoStop}%
\bibitem [{\citenamefont {Hohenberg}\ and\ \citenamefont {Halperin}(1977)}]{hohenberg1977theory}%
  \BibitemOpen
  \bibfield  {author} {\bibinfo {author} {\bibfnamefont {P.~C.}\ \bibnamefont {Hohenberg}}\ and\ \bibinfo {author} {\bibfnamefont {B.~I.}\ \bibnamefont {Halperin}},\ }\bibfield  {title} {\enquote {\bibinfo {title} {Theory of dynamic critical phenomena},}\ }\href@noop {} {\bibfield  {journal} {\bibinfo  {journal} {Reviews of Modern Physics}\ }\textbf {\bibinfo {volume} {49}},\ \bibinfo {pages} {435} (\bibinfo {year} {1977})}\BibitemShut {NoStop}%
\bibitem [{\citenamefont {Siggia}, \citenamefont {Halperin},\ and\ \citenamefont {Hohenberg}(1976)}]{siggia1976renormalization}%
  \BibitemOpen
  \bibfield  {author} {\bibinfo {author} {\bibfnamefont {E.}~\bibnamefont {Siggia}}, \bibinfo {author} {\bibfnamefont {B.}~\bibnamefont {Halperin}},\ and\ \bibinfo {author} {\bibfnamefont {P.}~\bibnamefont {Hohenberg}},\ }\bibfield  {title} {\enquote {\bibinfo {title} {Renormalization-group treatment of the critical dynamics of the binary-fluid and gas-liquid transitions},}\ }\href@noop {} {\bibfield  {journal} {\bibinfo  {journal} {Physical Review B}\ }\textbf {\bibinfo {volume} {13}},\ \bibinfo {pages} {2110} (\bibinfo {year} {1976})}\BibitemShut {NoStop}%
\bibitem [{\citenamefont {Chen}, \citenamefont {Tan},\ and\ \citenamefont {Fu}(2024)}]{chen2024critical}%
  \BibitemOpen
  \bibfield  {author} {\bibinfo {author} {\bibfnamefont {Y.-r.}\ \bibnamefont {Chen}}, \bibinfo {author} {\bibfnamefont {Y.-y.}\ \bibnamefont {Tan}},\ and\ \bibinfo {author} {\bibfnamefont {W.-j.}\ \bibnamefont {Fu}},\ }\bibfield  {title} {\enquote {\bibinfo {title} {Critical dynamics of model h within the real-time frg approach},}\ }\href@noop {} {\bibfield  {journal} {\bibinfo  {journal} {arXiv preprint arXiv:2406.00679}\ } (\bibinfo {year} {2024})}\BibitemShut {NoStop}%
\bibitem [{\citenamefont {Pitaevskii}\ and\ \citenamefont {Lifshitz}(2012)}]{pitaevskii2012physical}%
  \BibitemOpen
  \bibfield  {author} {\bibinfo {author} {\bibfnamefont {L.~P.}\ \bibnamefont {Pitaevskii}}\ and\ \bibinfo {author} {\bibfnamefont {E.}~\bibnamefont {Lifshitz}},\ }\href@noop {} {\emph {\bibinfo {title} {Physical Kinetics: Volume 10}}},\ Vol.~\bibinfo {volume} {10}\ (\bibinfo  {publisher} {Butterworth-Heinemann},\ \bibinfo {year} {2012})\BibitemShut {NoStop}%
\bibitem [{\citenamefont {Kawasaki}(1966)}]{kawasaki1966correlation}%
  \BibitemOpen
  \bibfield  {author} {\bibinfo {author} {\bibfnamefont {K.}~\bibnamefont {Kawasaki}},\ }\bibfield  {title} {\enquote {\bibinfo {title} {Correlation-function approach to the transport coefficients near the critical point. i},}\ }\href@noop {} {\bibfield  {journal} {\bibinfo  {journal} {Physical Review}\ }\textbf {\bibinfo {volume} {150}},\ \bibinfo {pages} {291} (\bibinfo {year} {1966})}\BibitemShut {NoStop}%
\bibitem [{\citenamefont {Stillinger}\ and\ \citenamefont {Weber}(1978)}]{stillinger1978study}%
  \BibitemOpen
  \bibfield  {author} {\bibinfo {author} {\bibfnamefont {F.~H.}\ \bibnamefont {Stillinger}}\ and\ \bibinfo {author} {\bibfnamefont {T.~A.}\ \bibnamefont {Weber}},\ }\bibfield  {title} {\enquote {\bibinfo {title} {Study of melting and freezing in the gaussian core model by molecular dynamics simulation},}\ }\href@noop {} {\bibfield  {journal} {\bibinfo  {journal} {The Journal of Chemical Physics}\ }\textbf {\bibinfo {volume} {68}},\ \bibinfo {pages} {3837--3844} (\bibinfo {year} {1978})}\BibitemShut {NoStop}%
\bibitem [{\citenamefont {Likos}(2001)}]{likos2001effective}%
  \BibitemOpen
  \bibfield  {author} {\bibinfo {author} {\bibfnamefont {C.~N.}\ \bibnamefont {Likos}},\ }\bibfield  {title} {\enquote {\bibinfo {title} {Effective interactions in soft condensed matter physics},}\ }\href@noop {} {\bibfield  {journal} {\bibinfo  {journal} {Physics Reports}\ }\textbf {\bibinfo {volume} {348}},\ \bibinfo {pages} {267--439} (\bibinfo {year} {2001})}\BibitemShut {NoStop}%
\bibitem [{\citenamefont {Budkov}(2019)}]{budkov2019statistical}%
  \BibitemOpen
  \bibfield  {author} {\bibinfo {author} {\bibfnamefont {Y.~A.}\ \bibnamefont {Budkov}},\ }\bibfield  {title} {\enquote {\bibinfo {title} {Statistical theory of fluids with a complex electric structure: Application to solutions of soft-core dipolar particles},}\ }\href@noop {} {\bibfield  {journal} {\bibinfo  {journal} {Fluid Phase Equilibria}\ }\textbf {\bibinfo {volume} {490}},\ \bibinfo {pages} {133--140} (\bibinfo {year} {2019})}\BibitemShut {NoStop}%
\bibitem [{\citenamefont {Kr{\"u}ger}\ \emph {et~al.}(2018)\citenamefont {Kr{\"u}ger}, \citenamefont {Solon}, \citenamefont {D{\'e}mery}, \citenamefont {Rohwer},\ and\ \citenamefont {Dean}}]{kruger2018stresses}%
  \BibitemOpen
  \bibfield  {author} {\bibinfo {author} {\bibfnamefont {M.}~\bibnamefont {Kr{\"u}ger}}, \bibinfo {author} {\bibfnamefont {A.}~\bibnamefont {Solon}}, \bibinfo {author} {\bibfnamefont {V.}~\bibnamefont {D{\'e}mery}}, \bibinfo {author} {\bibfnamefont {C.~M.}\ \bibnamefont {Rohwer}},\ and\ \bibinfo {author} {\bibfnamefont {D.~S.}\ \bibnamefont {Dean}},\ }\bibfield  {title} {\enquote {\bibinfo {title} {Stresses in non-equilibrium fluids: Exact formulation and coarse-grained theory},}\ }\href@noop {} {\bibfield  {journal} {\bibinfo  {journal} {The Journal of chemical physics}\ }\textbf {\bibinfo {volume} {148}} (\bibinfo {year} {2018})}\BibitemShut {NoStop}%
\bibitem [{\citenamefont {Brady}\ and\ \citenamefont {Morris}(1997)}]{brady1997microstructure}%
  \BibitemOpen
  \bibfield  {author} {\bibinfo {author} {\bibfnamefont {J.~F.}\ \bibnamefont {Brady}}\ and\ \bibinfo {author} {\bibfnamefont {J.~F.}\ \bibnamefont {Morris}},\ }\bibfield  {title} {\enquote {\bibinfo {title} {Microstructure of strongly sheared suspensions and its impact on rheology and diffusion},}\ }\href@noop {} {\bibfield  {journal} {\bibinfo  {journal} {Journal of Fluid Mechanics}\ }\textbf {\bibinfo {volume} {348}},\ \bibinfo {pages} {103--139} (\bibinfo {year} {1997})}\BibitemShut {NoStop}%
\bibitem [{\citenamefont {Russel}\ and\ \citenamefont {Saville}(1989)}]{russelschowalter}%
  \BibitemOpen
  \bibfield  {author} {\bibinfo {author} {\bibfnamefont {S.}~\bibnamefont {Russel}}\ and\ \bibinfo {author} {\bibfnamefont {D.}~\bibnamefont {Saville}},\ }\href@noop {} {\emph {\bibinfo {title} {Colloidal Dispersions}}}\ (\bibinfo  {publisher} {Cambridge: Cambridge University Press},\ \bibinfo {year} {1989})\BibitemShut {NoStop}%
\bibitem [{\citenamefont {Bird}, \citenamefont {Stewai},\ and\ \citenamefont {Lightfoot}(2002)}]{bird2002phenomena}%
  \BibitemOpen
  \bibfield  {author} {\bibinfo {author} {\bibfnamefont {R.~B.}\ \bibnamefont {Bird}}, \bibinfo {author} {\bibfnamefont {W.~E.}\ \bibnamefont {Stewai}},\ and\ \bibinfo {author} {\bibfnamefont {E.~N.}\ \bibnamefont {Lightfoot}},\ }\bibfield  {title} {\enquote {\bibinfo {title} {Transport phenomena},}\ }\href@noop {} {\  (\bibinfo {year} {2002})}\BibitemShut {NoStop}%
\bibitem [{\citenamefont {Landau}\ and\ \citenamefont {Lifshitz}(1987)}]{landau1987fluid}%
  \BibitemOpen
  \bibfield  {author} {\bibinfo {author} {\bibfnamefont {L.~D.}\ \bibnamefont {Landau}}\ and\ \bibinfo {author} {\bibfnamefont {E.~M.}\ \bibnamefont {Lifshitz}},\ }\href@noop {} {\emph {\bibinfo {title} {Fluid Mechanics: Volume 6}}},\ Vol.~\bibinfo {volume} {6}\ (\bibinfo  {publisher} {Elsevier},\ \bibinfo {year} {1987})\BibitemShut {NoStop}%
\bibitem [{\citenamefont {Wertheim}(1963)}]{wertheim1963exact}%
  \BibitemOpen
  \bibfield  {author} {\bibinfo {author} {\bibfnamefont {M.}~\bibnamefont {Wertheim}},\ }\bibfield  {title} {\enquote {\bibinfo {title} {Exact solution of the percus-yevick integral equation for hard spheres},}\ }\href@noop {} {\bibfield  {journal} {\bibinfo  {journal} {Physical Review Letters}\ }\textbf {\bibinfo {volume} {10}},\ \bibinfo {pages} {321} (\bibinfo {year} {1963})}\BibitemShut {NoStop}%
\bibitem [{\citenamefont {Thiele}(1963)}]{thiele1963equation}%
  \BibitemOpen
  \bibfield  {author} {\bibinfo {author} {\bibfnamefont {E.}~\bibnamefont {Thiele}},\ }\bibfield  {title} {\enquote {\bibinfo {title} {Equation of state for hard spheres},}\ }\href@noop {} {\bibfield  {journal} {\bibinfo  {journal} {Journal of Chemical Physics}\ }\textbf {\bibinfo {volume} {39}},\ \bibinfo {pages} {474--479} (\bibinfo {year} {1963})}\BibitemShut {NoStop}%
\bibitem [{\citenamefont {G{\'e}rard-Varet}\ and\ \citenamefont {Hillairet}(2020)}]{gerard2020analysis}%
  \BibitemOpen
  \bibfield  {author} {\bibinfo {author} {\bibfnamefont {D.}~\bibnamefont {G{\'e}rard-Varet}}\ and\ \bibinfo {author} {\bibfnamefont {M.}~\bibnamefont {Hillairet}},\ }\bibfield  {title} {\enquote {\bibinfo {title} {Analysis of the viscosity of dilute suspensions beyond einstein's formula},}\ }\href@noop {} {\bibfield  {journal} {\bibinfo  {journal} {Archive for Rational Mechanics and Analysis}\ }\textbf {\bibinfo {volume} {238}},\ \bibinfo {pages} {1349--1411} (\bibinfo {year} {2020})}\BibitemShut {NoStop}%
\bibitem [{\citenamefont {Levin}(2002)}]{levin2002electrostatic}%
  \BibitemOpen
  \bibfield  {author} {\bibinfo {author} {\bibfnamefont {Y.}~\bibnamefont {Levin}},\ }\bibfield  {title} {\enquote {\bibinfo {title} {Electrostatic correlations: from plasma to biology},}\ }\href@noop {} {\bibfield  {journal} {\bibinfo  {journal} {Reports on progress in physics}\ }\textbf {\bibinfo {volume} {65}},\ \bibinfo {pages} {1577} (\bibinfo {year} {2002})}\BibitemShut {NoStop}%
\bibitem [{\citenamefont {Brilliantov}(1998)}]{brilliantov1998accurate}%
  \BibitemOpen
  \bibfield  {author} {\bibinfo {author} {\bibfnamefont {N.~V.}\ \bibnamefont {Brilliantov}},\ }\bibfield  {title} {\enquote {\bibinfo {title} {Accurate first-principle equation of state for the one-component plasma},}\ }\href@noop {} {\bibfield  {journal} {\bibinfo  {journal} {Contributions to Plasma Physics}\ }\textbf {\bibinfo {volume} {38}},\ \bibinfo {pages} {489--499} (\bibinfo {year} {1998})}\BibitemShut {NoStop}%
\bibitem [{\citenamefont {Khrapak}\ and\ \citenamefont {Khrapak}(2016)}]{khrapak2016internal}%
  \BibitemOpen
  \bibfield  {author} {\bibinfo {author} {\bibfnamefont {S.~A.}\ \bibnamefont {Khrapak}}\ and\ \bibinfo {author} {\bibfnamefont {A.}~\bibnamefont {Khrapak}},\ }\bibfield  {title} {\enquote {\bibinfo {title} {Internal energy of the classical two-and three-dimensional one-component-plasma},}\ }\href@noop {} {\bibfield  {journal} {\bibinfo  {journal} {Contributions to Plasma Physics}\ }\textbf {\bibinfo {volume} {56}},\ \bibinfo {pages} {270--280} (\bibinfo {year} {2016})}\BibitemShut {NoStop}%
\bibitem [{\citenamefont {Budkov}\ \emph {et~al.}(2015)\citenamefont {Budkov}, \citenamefont {Kolesnikov}, \citenamefont {Georgi}, \citenamefont {Nogovitsyn},\ and\ \citenamefont {Kiselev}}]{budkov2015new}%
  \BibitemOpen
  \bibfield  {author} {\bibinfo {author} {\bibfnamefont {Y.~A.}\ \bibnamefont {Budkov}}, \bibinfo {author} {\bibfnamefont {A.}~\bibnamefont {Kolesnikov}}, \bibinfo {author} {\bibfnamefont {N.}~\bibnamefont {Georgi}}, \bibinfo {author} {\bibfnamefont {E.}~\bibnamefont {Nogovitsyn}},\ and\ \bibinfo {author} {\bibfnamefont {M.}~\bibnamefont {Kiselev}},\ }\bibfield  {title} {\enquote {\bibinfo {title} {A new equation of state of a flexible-chain polyelectrolyte solution: Phase equilibria and osmotic pressure in the salt-free case},}\ }\href@noop {} {\bibfield  {journal} {\bibinfo  {journal} {The Journal of Chemical Physics}\ }\textbf {\bibinfo {volume} {142}} (\bibinfo {year} {2015})}\BibitemShut {NoStop}%
\bibitem [{\citenamefont {Budkov}\ \emph {et~al.}(2013)\citenamefont {Budkov}, \citenamefont {Frolov}, \citenamefont {Kiselev},\ and\ \citenamefont {Brilliantov}}]{budkov2013surface}%
  \BibitemOpen
  \bibfield  {author} {\bibinfo {author} {\bibfnamefont {Y.~A.}\ \bibnamefont {Budkov}}, \bibinfo {author} {\bibfnamefont {A.}~\bibnamefont {Frolov}}, \bibinfo {author} {\bibfnamefont {M.}~\bibnamefont {Kiselev}},\ and\ \bibinfo {author} {\bibfnamefont {N.}~\bibnamefont {Brilliantov}},\ }\bibfield  {title} {\enquote {\bibinfo {title} {Surface-induced liquid-gas transition in salt-free solutions of model charged colloids},}\ }\href@noop {} {\bibfield  {journal} {\bibinfo  {journal} {The Journal of Chemical Physics}\ }\textbf {\bibinfo {volume} {139}} (\bibinfo {year} {2013})}\BibitemShut {NoStop}%
\bibitem [{\citenamefont {Brilliantov}, \citenamefont {Kuznetsov},\ and\ \citenamefont {Klein}(1998)}]{brilliantov1998chain}%
  \BibitemOpen
  \bibfield  {author} {\bibinfo {author} {\bibfnamefont {N.}~\bibnamefont {Brilliantov}}, \bibinfo {author} {\bibfnamefont {D.}~\bibnamefont {Kuznetsov}},\ and\ \bibinfo {author} {\bibfnamefont {R.}~\bibnamefont {Klein}},\ }\bibfield  {title} {\enquote {\bibinfo {title} {Chain collapse and counterion condensation in dilute polyelectrolyte solutions},}\ }\href@noop {} {\bibfield  {journal} {\bibinfo  {journal} {Physical review letters}\ }\textbf {\bibinfo {volume} {81}},\ \bibinfo {pages} {1433} (\bibinfo {year} {1998})}\BibitemShut {NoStop}%
\bibitem [{\citenamefont {Falkenhagen}\ and\ \citenamefont {Vernon}(1932)}]{falkenhagen1932lxii}%
  \BibitemOpen
  \bibfield  {author} {\bibinfo {author} {\bibfnamefont {H.}~\bibnamefont {Falkenhagen}}\ and\ \bibinfo {author} {\bibfnamefont {E.}~\bibnamefont {Vernon}},\ }\bibfield  {title} {\enquote {\bibinfo {title} {Lxii. the viscosity of strong electrolyte solutions according to electrostatic theory},}\ }\href@noop {} {\bibfield  {journal} {\bibinfo  {journal} {The London, Edinburgh, and Dublin Philosophical Magazine and Journal of Science}\ }\textbf {\bibinfo {volume} {14}},\ \bibinfo {pages} {537--565} (\bibinfo {year} {1932})}\BibitemShut {NoStop}%
\bibitem [{\citenamefont {Onsager}\ and\ \citenamefont {Fuoss}(1932)}]{onsager2002irreversible}%
  \BibitemOpen
  \bibfield  {author} {\bibinfo {author} {\bibfnamefont {L.}~\bibnamefont {Onsager}}\ and\ \bibinfo {author} {\bibfnamefont {R.~M.}\ \bibnamefont {Fuoss}},\ }\bibfield  {title} {\enquote {\bibinfo {title} {Irreversible processes in electrolytes. diffusion, conductance and viscous flow in arbitrary mixtures of strong electrolytes},}\ }\href@noop {} {\bibfield  {journal} {\bibinfo  {journal} {The Journal of Physical Chemistry}\ }\textbf {\bibinfo {volume} {36}},\ \bibinfo {pages} {2689--2778} (\bibinfo {year} {1932})}\BibitemShut {NoStop}%
\bibitem [{\citenamefont {Fixman}(1962)}]{fixman1962viscosity}%
  \BibitemOpen
  \bibfield  {author} {\bibinfo {author} {\bibfnamefont {M.}~\bibnamefont {Fixman}},\ }\bibfield  {title} {\enquote {\bibinfo {title} {Viscosity of critical mixtures},}\ }\href@noop {} {\bibfield  {journal} {\bibinfo  {journal} {The Journal of Chemical Physics}\ }\textbf {\bibinfo {volume} {36}},\ \bibinfo {pages} {310--318} (\bibinfo {year} {1962})}\BibitemShut {NoStop}%
\bibitem [{\citenamefont {Gitterman}(1988)}]{gitterman1988dilute}%
  \BibitemOpen
  \bibfield  {author} {\bibinfo {author} {\bibfnamefont {M.}~\bibnamefont {Gitterman}},\ }\bibfield  {title} {\enquote {\bibinfo {title} {Are dilute solutions always dilute?}}\ }\href@noop {} {\bibfield  {journal} {\bibinfo  {journal} {American Journal of Physics}\ }\textbf {\bibinfo {volume} {56}},\ \bibinfo {pages} {1000--1002} (\bibinfo {year} {1988})}\BibitemShut {NoStop}%
\bibitem [{\citenamefont {Rotne}\ and\ \citenamefont {Prager}(1969)}]{rotne1969variational}%
  \BibitemOpen
  \bibfield  {author} {\bibinfo {author} {\bibfnamefont {J.}~\bibnamefont {Rotne}}\ and\ \bibinfo {author} {\bibfnamefont {S.}~\bibnamefont {Prager}},\ }\bibfield  {title} {\enquote {\bibinfo {title} {Variational treatment of hydrodynamic interaction in polymers},}\ }\href@noop {} {\bibfield  {journal} {\bibinfo  {journal} {The Journal of Chemical Physics}\ }\textbf {\bibinfo {volume} {50}},\ \bibinfo {pages} {4831--4837} (\bibinfo {year} {1969})}\BibitemShut {NoStop}%
\bibitem [{\citenamefont {Henrich}, \citenamefont {Pfeifroth},\ and\ \citenamefont {Fuchs}(2007)}]{henrich2007nonequilibrium}%
  \BibitemOpen
  \bibfield  {author} {\bibinfo {author} {\bibfnamefont {O.}~\bibnamefont {Henrich}}, \bibinfo {author} {\bibfnamefont {O.}~\bibnamefont {Pfeifroth}},\ and\ \bibinfo {author} {\bibfnamefont {M.}~\bibnamefont {Fuchs}},\ }\bibfield  {title} {\enquote {\bibinfo {title} {Nonequilibrium structure of concentrated colloidal fluids under steady shear: leading-orderresponse},}\ }\href@noop {} {\bibfield  {journal} {\bibinfo  {journal} {Journal of Physics: Condensed Matter}\ }\textbf {\bibinfo {volume} {19}},\ \bibinfo {pages} {205132} (\bibinfo {year} {2007})}\BibitemShut {NoStop}%
\bibitem [{\citenamefont {Bergenholtz}(2001)}]{bergenholtz2001theory}%
  \BibitemOpen
  \bibfield  {author} {\bibinfo {author} {\bibfnamefont {J.}~\bibnamefont {Bergenholtz}},\ }\bibfield  {title} {\enquote {\bibinfo {title} {Theory of rheology of colloidal dispersions},}\ }\href@noop {} {\bibfield  {journal} {\bibinfo  {journal} {Current opinion in colloid \& interface science}\ }\textbf {\bibinfo {volume} {6}},\ \bibinfo {pages} {484--488} (\bibinfo {year} {2001})}\BibitemShut {NoStop}%
\bibitem [{\citenamefont {Bernard}\ \emph {et~al.}(2023)\citenamefont {Bernard}, \citenamefont {Jardat}, \citenamefont {Rotenberg},\ and\ \citenamefont {Illien}}]{bernard2023analytical}%
  \BibitemOpen
  \bibfield  {author} {\bibinfo {author} {\bibfnamefont {O.}~\bibnamefont {Bernard}}, \bibinfo {author} {\bibfnamefont {M.}~\bibnamefont {Jardat}}, \bibinfo {author} {\bibfnamefont {B.}~\bibnamefont {Rotenberg}},\ and\ \bibinfo {author} {\bibfnamefont {P.}~\bibnamefont {Illien}},\ }\bibfield  {title} {\enquote {\bibinfo {title} {On analytical theories for conductivity and self-diffusion in concentrated electrolytes},}\ }\href@noop {} {\bibfield  {journal} {\bibinfo  {journal} {The Journal of Chemical Physics}\ }\textbf {\bibinfo {volume} {159}} (\bibinfo {year} {2023})}\BibitemShut {NoStop}%
\bibitem [{Note1()}]{Note1}%
  \BibitemOpen
  \bibinfo {note} {For the sake of simplicity, we assume that there is no external magnetic field present, for which the Onsager kinetic coefficients are always symmetrical.}\BibitemShut {Stop}%
\bibitem [{Note2()}]{Note2}%
  \BibitemOpen
  \bibinfo {note} {Only in the case where we can ignore the loss of particles due to vaporization at the free boundary.}\BibitemShut {Stop}%
\bibitem [{\citenamefont {Brilliantov}\ and\ \citenamefont {Revokatov}(1996)}]{brilliantov1996molekulyarnaya}%
  \BibitemOpen
  \bibfield  {author} {\bibinfo {author} {\bibfnamefont {N.}~\bibnamefont {Brilliantov}}\ and\ \bibinfo {author} {\bibfnamefont {O.}~\bibnamefont {Revokatov}},\ }\href@noop {} {\emph {\bibinfo {title} {Molecular dynamics of disordered media (in russian)}}}\ (\bibinfo  {publisher} {Moscow: Izdatel'stvo Moskovskogo universiteta},\ \bibinfo {year} {1996})\BibitemShut {NoStop}%
\bibitem [{\citenamefont {Landau}(1937)}]{landau1937towards}%
  \BibitemOpen
  \bibfield  {author} {\bibinfo {author} {\bibfnamefont {L.}~\bibnamefont {Landau}},\ }\bibfield  {title} {\enquote {\bibinfo {title} {Towards the theory of phase transitions. ii},}\ }\href@noop {} {\bibfield  {journal} {\bibinfo  {journal} {Phys. Z. Sowjetunion}\ }\textbf {\bibinfo {volume} {11}},\ \bibinfo {pages} {545} (\bibinfo {year} {1937})}\BibitemShut {NoStop}%
\bibitem [{\citenamefont {Kats}, \citenamefont {Lebedev},\ and\ \citenamefont {Muratov}(1993)}]{kats1993weak}%
  \BibitemOpen
  \bibfield  {author} {\bibinfo {author} {\bibfnamefont {E.}~\bibnamefont {Kats}}, \bibinfo {author} {\bibfnamefont {V.}~\bibnamefont {Lebedev}},\ and\ \bibinfo {author} {\bibfnamefont {A.}~\bibnamefont {Muratov}},\ }\bibfield  {title} {\enquote {\bibinfo {title} {Weak crystallization theory},}\ }\href@noop {} {\bibfield  {journal} {\bibinfo  {journal} {Physics reports}\ }\textbf {\bibinfo {volume} {228}},\ \bibinfo {pages} {1--91} (\bibinfo {year} {1993})}\BibitemShut {NoStop}%
\bibitem [{\citenamefont {Borue}\ and\ \citenamefont {Erukhimovich}(1988)}]{borue1988statistical}%
  \BibitemOpen
  \bibfield  {author} {\bibinfo {author} {\bibfnamefont {V.~Y.}\ \bibnamefont {Borue}}\ and\ \bibinfo {author} {\bibfnamefont {I.~Y.}\ \bibnamefont {Erukhimovich}},\ }\bibfield  {title} {\enquote {\bibinfo {title} {A statistical theory of weakly charged polyelectrolytes: fluctuations, equation of state and microphase separation},}\ }\href@noop {} {\bibfield  {journal} {\bibinfo  {journal} {Macromolecules}\ }\textbf {\bibinfo {volume} {21}},\ \bibinfo {pages} {3240--3249} (\bibinfo {year} {1988})}\BibitemShut {NoStop}%
\bibitem [{\citenamefont {De~La~Cruz}(1991)}]{de1991transitions}%
  \BibitemOpen
  \bibfield  {author} {\bibinfo {author} {\bibfnamefont {M.~O.}\ \bibnamefont {De~La~Cruz}},\ }\bibfield  {title} {\enquote {\bibinfo {title} {Transitions to periodic structures in block copolymer melts},}\ }\href@noop {} {\bibfield  {journal} {\bibinfo  {journal} {Physical review letters}\ }\textbf {\bibinfo {volume} {67}},\ \bibinfo {pages} {85} (\bibinfo {year} {1991})}\BibitemShut {NoStop}%
\bibitem [{\citenamefont {Khokhlov}\ and\ \citenamefont {Nyrkova}(1992)}]{khokhlov1992compatibility}%
  \BibitemOpen
  \bibfield  {author} {\bibinfo {author} {\bibfnamefont {A.}~\bibnamefont {Khokhlov}}\ and\ \bibinfo {author} {\bibfnamefont {I.}~\bibnamefont {Nyrkova}},\ }\bibfield  {title} {\enquote {\bibinfo {title} {Compatibility enhancement and microdomain structuring in weakly charged polyelectrolyte mixtures},}\ }\href@noop {} {\bibfield  {journal} {\bibinfo  {journal} {Macromolecules}\ }\textbf {\bibinfo {volume} {25}},\ \bibinfo {pages} {1493--1502} (\bibinfo {year} {1992})}\BibitemShut {NoStop}%
\bibitem [{\citenamefont {Archer}\ and\ \citenamefont {Evans}(2004)}]{archer2004dynamical}%
  \BibitemOpen
  \bibfield  {author} {\bibinfo {author} {\bibfnamefont {A.~J.}\ \bibnamefont {Archer}}\ and\ \bibinfo {author} {\bibfnamefont {R.}~\bibnamefont {Evans}},\ }\bibfield  {title} {\enquote {\bibinfo {title} {Dynamical density functional theory and its application to spinodal decomposition},}\ }\href@noop {} {\bibfield  {journal} {\bibinfo  {journal} {The Journal of chemical physics}\ }\textbf {\bibinfo {volume} {121}},\ \bibinfo {pages} {4246--4254} (\bibinfo {year} {2004})}\BibitemShut {NoStop}%
\bibitem [{\citenamefont {Nabutovskii}, \citenamefont {Nemov},\ and\ \citenamefont {Peisakhovich}(1980)}]{nabutovskii1980order}%
  \BibitemOpen
  \bibfield  {author} {\bibinfo {author} {\bibfnamefont {V.}~\bibnamefont {Nabutovskii}}, \bibinfo {author} {\bibfnamefont {N.}~\bibnamefont {Nemov}},\ and\ \bibinfo {author} {\bibfnamefont {Y.~G.}\ \bibnamefont {Peisakhovich}},\ }\bibfield  {title} {\enquote {\bibinfo {title} {Order-parameter and charge-density waves near the critical point in an electrolyte},}\ }\href@noop {} {\bibfield  {journal} {\bibinfo  {journal} {Soviet Journal of Experimental and Theoretical Physics}\ }\textbf {\bibinfo {volume} {52}},\ \bibinfo {pages} {1111} (\bibinfo {year} {1980})}\BibitemShut {NoStop}%
\bibitem [{\citenamefont {H{\o}ye}\ and\ \citenamefont {Stell}(1990)}]{hoye1990effect}%
  \BibitemOpen
  \bibfield  {author} {\bibinfo {author} {\bibfnamefont {J.}~\bibnamefont {H{\o}ye}}\ and\ \bibinfo {author} {\bibfnamefont {G.}~\bibnamefont {Stell}},\ }\bibfield  {title} {\enquote {\bibinfo {title} {Effect of solute on solvent critical behavior. 1. general formalism and the possibility of charge-density waves},}\ }\href@noop {} {\bibfield  {journal} {\bibinfo  {journal} {Journal of Physical Chemistry}\ }\textbf {\bibinfo {volume} {94}},\ \bibinfo {pages} {7899--7907} (\bibinfo {year} {1990})}\BibitemShut {NoStop}%
\bibitem [{\citenamefont {Cook}(1970)}]{cook1970brownian}%
  \BibitemOpen
  \bibfield  {author} {\bibinfo {author} {\bibfnamefont {H.}~\bibnamefont {Cook}},\ }\bibfield  {title} {\enquote {\bibinfo {title} {Brownian motion in spinodal decomposition},}\ }\href@noop {} {\bibfield  {journal} {\bibinfo  {journal} {Acta metallurgica}\ }\textbf {\bibinfo {volume} {18}},\ \bibinfo {pages} {297--306} (\bibinfo {year} {1970})}\BibitemShut {NoStop}%
\bibitem [{\citenamefont {Langer}(1971)}]{langer1971theory}%
  \BibitemOpen
  \bibfield  {author} {\bibinfo {author} {\bibfnamefont {J.~S.}\ \bibnamefont {Langer}},\ }\bibfield  {title} {\enquote {\bibinfo {title} {Theory of spinodal decomposition in alloys},}\ }\href@noop {} {\bibfield  {journal} {\bibinfo  {journal} {Annals of Physics}\ }\textbf {\bibinfo {volume} {65}},\ \bibinfo {pages} {53--86} (\bibinfo {year} {1971})}\BibitemShut {NoStop}%
\bibitem [{\citenamefont {Zakharov}(1978)}]{zakharov1978impurity}%
  \BibitemOpen
  \bibfield  {author} {\bibinfo {author} {\bibfnamefont {A.~Y.}\ \bibnamefont {Zakharov}},\ }\bibfield  {title} {\enquote {\bibinfo {title} {Impurity diffusion and distribution in alloys},}\ }\href@noop {} {\bibfield  {journal} {\bibinfo  {journal} {Solid State Communications}\ }\textbf {\bibinfo {volume} {28}},\ \bibinfo {pages} {811--813} (\bibinfo {year} {1978})}\BibitemShut {NoStop}%
\bibitem [{\citenamefont {Skripov}\ and\ \citenamefont {Skripov}(1979)}]{skripov1979spinodal}%
  \BibitemOpen
  \bibfield  {author} {\bibinfo {author} {\bibfnamefont {V.~P.}\ \bibnamefont {Skripov}}\ and\ \bibinfo {author} {\bibfnamefont {A.}~\bibnamefont {Skripov}},\ }\bibfield  {title} {\enquote {\bibinfo {title} {Spinodal decomposition (phase transitions via unstable states)},}\ }\href@noop {} {\bibfield  {journal} {\bibinfo  {journal} {Soviet Physics Uspekhi}\ }\textbf {\bibinfo {volume} {22}},\ \bibinfo {pages} {389} (\bibinfo {year} {1979})}\BibitemShut {NoStop}%
\bibitem [{\citenamefont {Brazovskii}, \citenamefont {Dzyaloshinskii},\ and\ \citenamefont {Muratov}(1987)}]{brazovskii1987theory}%
  \BibitemOpen
  \bibfield  {author} {\bibinfo {author} {\bibfnamefont {S.}~\bibnamefont {Brazovskii}}, \bibinfo {author} {\bibfnamefont {I.}~\bibnamefont {Dzyaloshinskii}},\ and\ \bibinfo {author} {\bibfnamefont {A.}~\bibnamefont {Muratov}},\ }\bibfield  {title} {\enquote {\bibinfo {title} {Theory of weak crystallization},}\ }\href@noop {} {\bibfield  {journal} {\bibinfo  {journal} {Sov. Phys. JETP}\ }\textbf {\bibinfo {volume} {66}},\ \bibinfo {pages} {625} (\bibinfo {year} {1987})}\BibitemShut {NoStop}%
\bibitem [{\citenamefont {Erukhimovich}\ and\ \citenamefont {Kriksin}(2019)}]{erukhimovich2019thermodynamics}%
  \BibitemOpen
  \bibfield  {author} {\bibinfo {author} {\bibfnamefont {I.}~\bibnamefont {Erukhimovich}}\ and\ \bibinfo {author} {\bibfnamefont {Y.}~\bibnamefont {Kriksin}},\ }\bibfield  {title} {\enquote {\bibinfo {title} {Thermodynamics of 3d diamond-like epitaxial (film) morphologies on 1d modulated substrate: Weak crystallization theory},}\ }\href@noop {} {\bibfield  {journal} {\bibinfo  {journal} {The Journal of Chemical Physics}\ }\textbf {\bibinfo {volume} {150}} (\bibinfo {year} {2019})}\BibitemShut {NoStop}%
\end{thebibliography}%
\end{document}